\documentclass[apj]{emulateapj}

\slugcomment{}

\shorttitle{Ensemble Variability of near-infrared AGNs}
\shortauthors{Kouzuma \& Yamaoka}

\begin{document}
\title{Ensemble Variability of Near-infrared Selected Active Galactic Nuclei}

\author{S. Kouzuma}
\affil{School of International Liberal Studies, Chukyo University, Toyota 470-0393, Japan}
\email{skouzuma@lets.chukyo-u.ac.jp}
\and
\author{H. Yamaoka}
\affil{Department of Physics, Kyushu University, Fukuoka 812-8581, Japan}
\email{yamaoka@phys.kyushu-u.ac.jp}

\begin{abstract}
We present the properties of the ensemble variability $V$ for nearly 5,000 near-infrared active galactic nuclei (AGNs) selected from 
the catalog of Quasars and Active Galactic Nuclei (13th Edition) and the SDSS-DR7 quasar catalog. 
From three near-infrared point source catalogs; namely 2MASS, DENIS, and UKIDSS/LAS catalogs, 
we extract 2MASS-DENIS and 2MASS-UKIDSS counterparts for cataloged AGNs by cross-identification between catalogs. 
We further select variable AGNs based on an optimal criterion for selecting the variable sources. 
The sample objects are divided into subsets according to whether near-infrared light 
originates by optical emission or by near-infrared emission, in the rest-frame; and 
we examine the correlations of the ensemble variability with the rest-frame wavelength, redshift, luminosity, and rest-frame time lag. 
In addition, we also examine the correlations of variability amplitude with optical variability, radio intensity, and radio-to-optical flux ratio. 
The rest-frame optical variability of our samples shows negative correlations with luminosity and positive correlations with rest-frame time lag 
(i.e., the structure function, SF), 
and this result is consistent with previous analyses. 
However, no well-known negative correlation exists between the rest-frame wavelength and optical variability. 
This inconsistency might be due to a biased sampling of high-redshift AGNs. 
Near-infrared variability in the rest-frame is anticorrelated with the rest-frame wavelength, 
which is consistent with previous suggestions. 
However, correlations of near-infrared variability with luminosity and rest-frame time lag 
are the opposite of these correlations of the optical variability; 
that is, the near-infrared variability is positively correlated with luminosity but negatively correlated with the rest-frame time lag. 
Because these trends are qualitatively consistent with the properties of radio-loud quasars reported by some previous studies, 
most of our sample objects are probably radio-loud quasars. 
Finally, we also discuss the negative correlations seen in the near-infrared SFs. 
\end{abstract}

\keywords{galaxies: active -- quasars: general -- catalogs}

\section{Introduction}
Variability is a characteristic property of active galactic nuclei (AGNs). 
It is believed that the majority of AGNs exhibit rapid and apparently random variability over a wide wavelength band 
that ranges from x-ray to radio wavelengths. 
Characteristic variability timescales range from months to years. 
Variability is a powerful feature for constraining physical models and 
has the potential to probe the structure and dynamics of the central engine. 

It is widely accepted that accretion disk instabilities represent a promising model 
for explaining the variability properties of AGNs \citep[e.g.,][]{Kawaguchi1998-ApJ}. 
In addition, several possible models have been proposed to explain the observed optical variability: 
multiple supernovae \citep{Terlevich1992-MNRAS} or starburst \citep{Terlevich1992-MNRAS,Aretxaga1997-MNRAS} 
close to the nucleus, 
star collisions \citep{Courvoisier1996-AA,Torricelli2000-AA}, and 
gravitational microlensing \citep{Hawkins1993-Nature,Zackrisson2003-AA}. 
A powerful approach for discerning these models is to examine correlations between variability and a physical parameter 
(such as time lag, luminosity, rest-frame wavelength, or redshift). 
These observational correlations have been associated with correlations predicted by physical models, and 
the appropriate model for the variability mechanism of AGNs has often been discussed. 

There are two main approaches to examine the variability of AGNs. 
The first is to monitor a relatively small sample of AGNs over a period of time 
\citep[e.g.,][]{Garcia1999-MNRAS,Neugebauer1999-AJ,Walsh2009-ApJS}. 
By using an observed light curve for an AGN, various variability properties for the object were studied. 
The second approach is to examine the ensemble variability $V$ of a larger number of AGNs \citep[e.g.,][]{Vries2003-AJ,Bauer2009-ApJ}. 
This approach can be accomplished even if there are only two sampling epochs. 
For instance, \citet{VandenBerk2004-ApJ} studied the ensemble variability of 25,000 quasars in the SDSS catalog using two sampling epochs 
and established several correlations between the optical variability and time lag, luminosity, rest-frame wavelength, and redshift. 
In the near-infrared, \citet{Neugebauer1989-AJ} studied the near-infrared variability of individual quasars 
using a sample of 108 optically selected quasars. 
In addition, near-infrared ensemble variability is reported by \citet{Enya2002-ApJS}. 
They presented correlations between near-infrared variability and several physical parameters by studying a sample of 226 AGNs. 
Whereas numerous studies have been dedicated to examining the optical variability 
(especially to establish a correlation between optical variability and a physical parameter), 
a few studies have focused on the near-infrared variability. 

In this paper, we present the optical and near-infrared ensemble variability in the rest-frame 
for a sample of about 5,000 AGNs, which are probably mostly quasars, 
by using archival three near-infrared catalogs and two AGN catalogs. 
The catalogs used in this paper are introduced in Section \ref{Data}. 
In Section \ref{Selection}, we describe the method for extracting near-infrared counterparts for cataloged AGNs and 
the criterion for selecting near-infrared variable AGNs. 
In Section \ref{Analysis}, we show correlations of the ensemble variability with rest-frame wavelength, 
redshift, luminosity, and rest-frame time lag. 
We also show the variability properties correlated with optical variability, radio intensity, and radio-to-optical flux ratio. 
Finally, we discuss both the intrinsic optical and near-infrared variability properties in Section \ref{Discussion}.

%%%%%%%%%%%%%%%%%%%%%%%%%%%%%%%%%%%%%%%%%%%%%%%%%%
\section{Data}\label{Data}
\subsection{2MASS}
The Two Micron All Sky Survey (2MASS)\footnote{2MASS web site (http://www.ipac.caltech.edu/2mass/)} \citep{Skrutskie2006-AJ}
is a project that observed 99.998\% of the whole sky 
at the J (1.25 $\mu$m), H (1.65 $\mu$m), and Ks (2.16 $\mu$m) bands, 
from Mt. Hopkins, Arizona, USA (in the northern hemisphere) and from CTIO, Chile (in the southern hemisphere)
between June, 1997 and February, 2001. 
Both the instruments are highly automated 1.3-m telescopes equipped with three-channel cameras, 
with each channel consisting of a 256 $\times$ 256 array of HgCdTe detectors. 
2MASS obtained 4,121,439 FITS images (pixel size $\sim2''_{\cdot}0$) with an integration time of 7.8 s.
The limiting magnitudes [signal-to-noise ratio $S/N$$>$10] are 
15.8 (J), 15.1 (H), and 14.3 (K$_\textnormal{\tiny S}$) mag at each band.
The Point Source Catalog (PSC) was produced using these images and contains 470,992,970 sources.
The images and the PSC are publicly available on the 2MASS web site.

The 2MASS default magnitude in the PSC is measured by a point spread function (PSF) profile-fitting algorithm. 
If no source is detected, this is the 95\% confidence upper limit 
derived from a $4''$ radius aperture measurement taken at the position of the source on the Atlas Image.
The uncertainties in the profile-fit photometry take into account the PSF shape, using a measurement-noise model that includes 
the effects of read noise, Poisson noise, and PSF error.

%%%%%%%%%%%%%%%%%%%%%%%%%%%%%%
\subsection{DENIS}
The Deep Near Infrared Survey of the Southern sky (DENIS)\footnote{http://cdsweb.u-strasbg.fr/denis.html} 
is a survey of the southern sky done between 1995 and 2001 in two near-infrared bands 
(J: $1.25$ $\mu$m and K$_\textnormal{\tiny S}$: $2.15$ $\mu$m) and one optical band (I: $0.82$ $\mu$m). 
The instrument is a 1-m ground-based telescope at La Silla (Chile) 
with limiting magnitudes of 16.5, 14.0, and 18.5 in the three bands, respectively. 
The positional accuracy is better than $1''$. 
The DENIS catalog consists of a set of $\sim 350$ million point sources in approximately 16,700 deg$^2$ of the southern sky. 
The PSF and several kinds of aperture magnitudes are available in the DENIS catalog. 
The photometric quality is defined by quality flags, which are represented by a number between 0 and 100.

%%%%%%%%%%%%%%%%%%%%%%%%%%%%%
\subsection{UKIDSS}
The United Kingdom Infrared Telescope (UKIRT) Infrared Deep Sky Survey (UKIDSS)\footnote{http://www.ukidss.org/} \citep{Lawrence2007-MNRAS} 
began in May 2005 and was planned to survey 7,500 deg$^2$ of the northern sky, 
extending over both high and low galactic latitudes. 
The survey instrument is the UKIRT Wide Field Camera \citep[WFCAM;][]{Casali2007-AA}, 
which has four $2048\times 2048$ Rockwell devices, on the UKIRT in Hawaii. 
The pixel scale is $0''_{\cdot}4$. 
The pipeline processing and science archive are described in Irwin et al (2008) and Hambly et al (2008). 

The UKIDSS consists of five surveys: 
the Large Area Survey (LAS), 
Galactic Plane Survey (GPS), Galactic Clusters Survey (GCS), 
Deep Extragalactic Survey (DXS), and Ultra Deep Survey (UDS). 
In this paper, we use the LAS data from UKIDSS DR6.
The UKIDSS/LAS covers an area of $\sim$4,000 deg$^2$ at high galactic latitude in the YJHK filters. 
The photometric system is described in \citet{Hewett2006-MNRAS}. 

The UKIDSS/LAS point source catalog contains the aperture magnitude measured using several aperture sizes. 
The photometric errors are less than 0.02 mag in all filters, and 
the global uniformity is as good as the 2MASS survey \citep{Lawrence2007-MNRAS}.

%%%%%%%%%%%%%%%%%%%%%%%%%%%%%
\subsection{Sample AGNs}
Sample AGNs are extracted from two catalogs: 
the catalog of Quasars and Active Galactic Nuclei 
\citep[13th Ed;][hereafter QA]{Veron2010-AA} and the SDSS-DR7 quasar catalog \citep[hereafter SQ]{Schneider2010-AJ}. 
The QA catalog includes 133,336 quasars, 1,374 BL Lac objects, and 34,231 active galaxies 
(including 15,627 Seyfert 1) with positional accuracy superior to $1''_{\cdot}0$. 
The SQ catalog includes 105,783 spectroscopically confirmed quasars in an area covering $\sim$9,380 deg$^2$, 
and has a positional accuracy better than $0''_{\cdot}1$ rms per coordinate.

\begin{table}[thbp]
\begin{center}
\caption{Number of cataloged AGNs having near-infrared counterparts. 
\label{Counterparts-Number}}
\begin{tabular}{lrrrr}
\hline 
 & \multicolumn{2}{c}{DENIS} & \multicolumn{2}{c}{UKIDSS} \\
 & \multicolumn{1}{c}{J} & \multicolumn{1}{c}{K} & \multicolumn{1}{c}{J} & \multicolumn{1}{c}{K} \\ \hline
QA & 2750 & 1159 & 3832 & 3540 \\
SQ &  551 &  110 & 2484 & 1945 \\ \hline
\end{tabular}
\end{center}
\end{table}

\begin{figure*}[thbp]
	\begin{center}
		\resizebox{80mm}{!}{\includegraphics[clip]{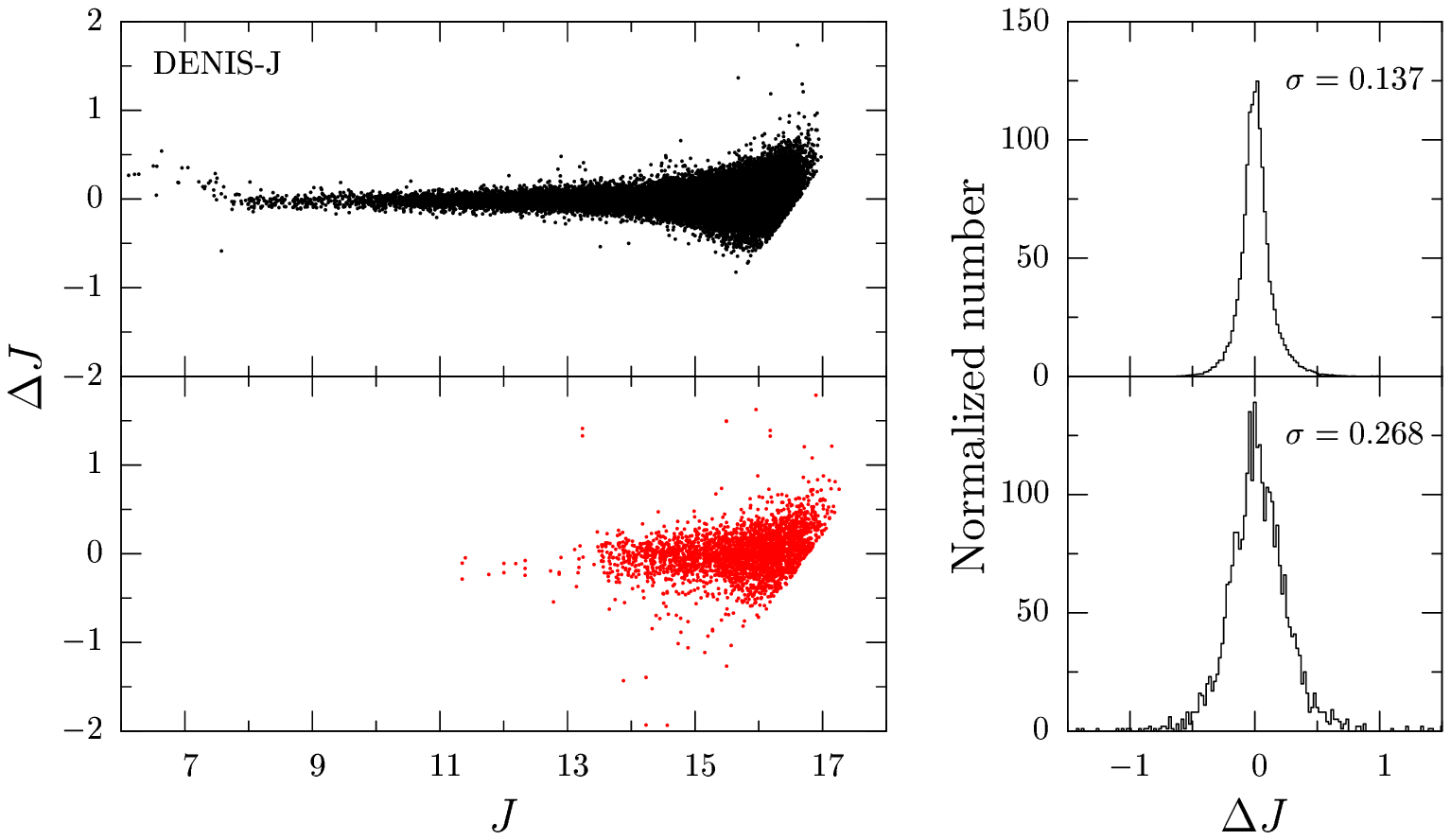}}
		\resizebox{80mm}{!}{\includegraphics[clip]{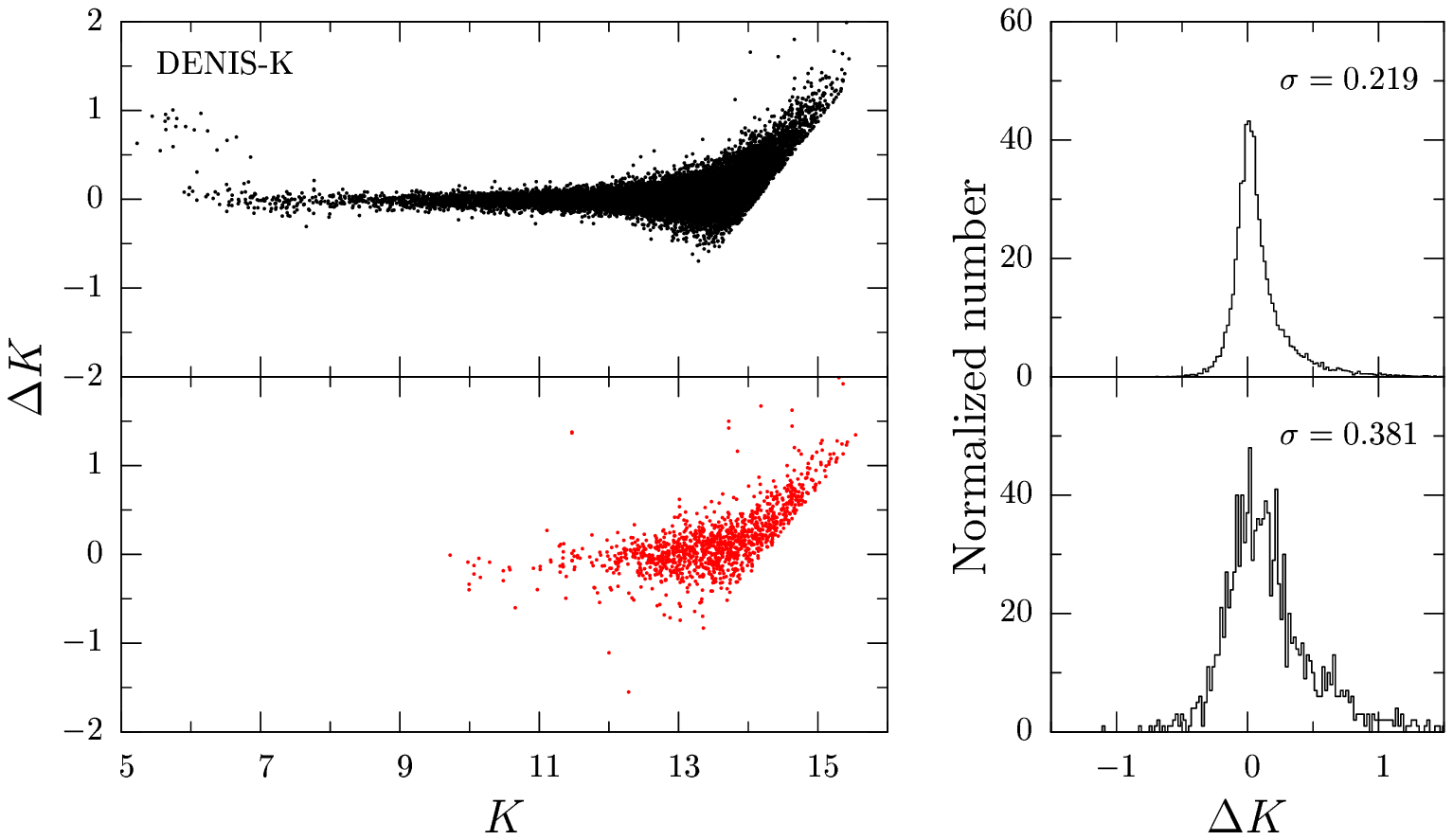}}
		\resizebox{80mm}{!}{\includegraphics[clip]{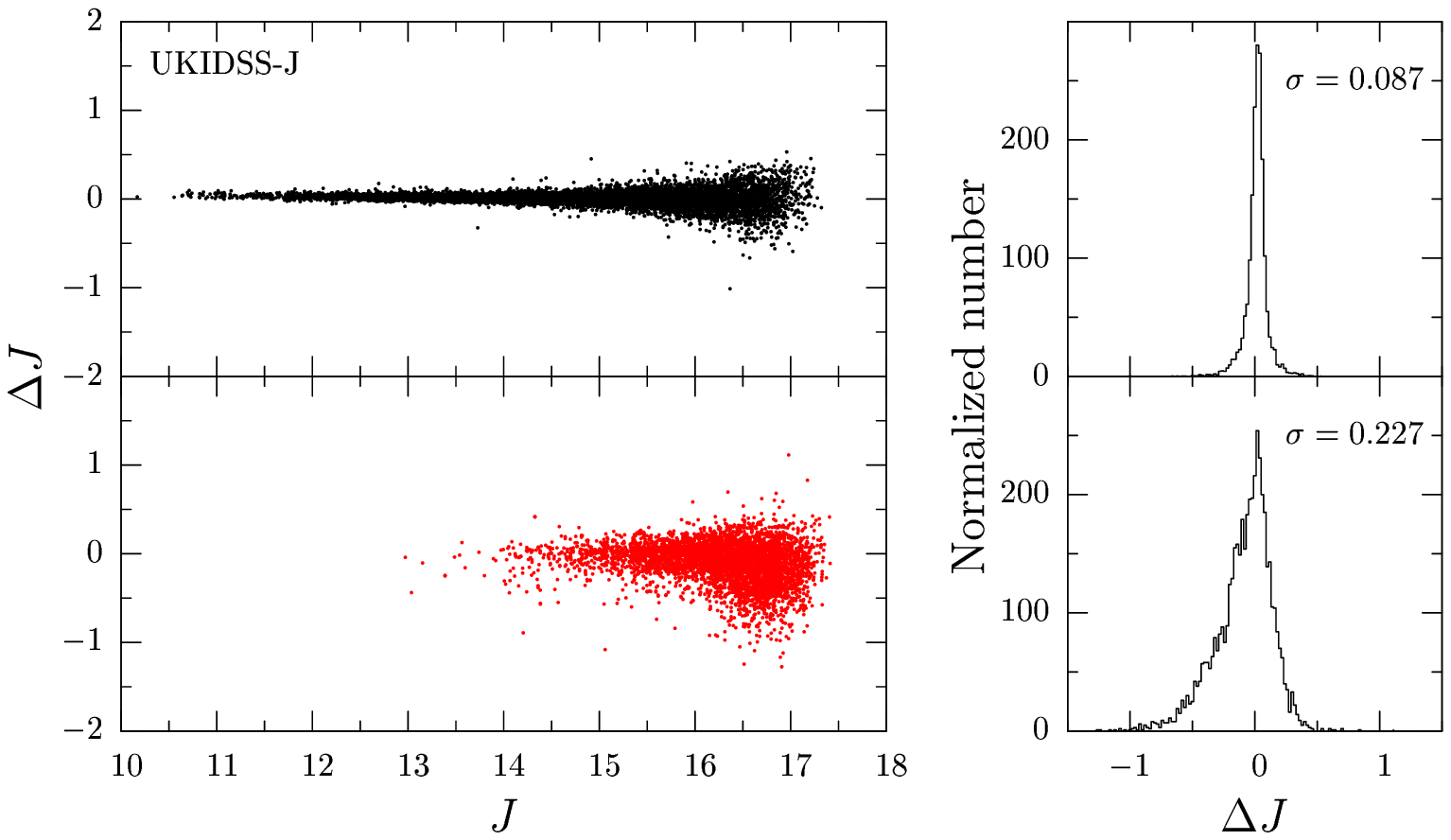}}
		\resizebox{80mm}{!}{\includegraphics[clip]{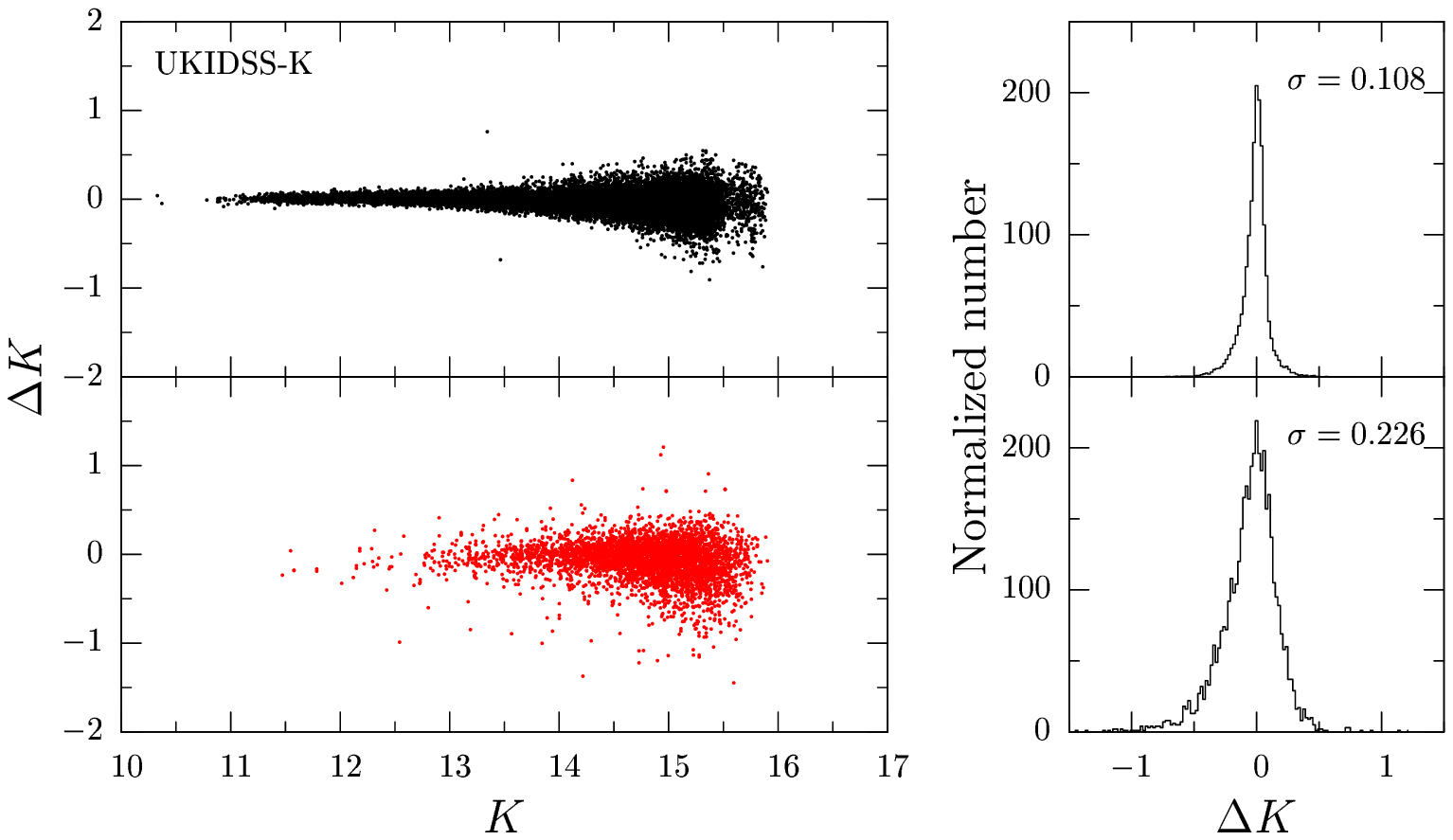}}
	\end{center}
		\caption{Magnitude versus magnitude difference between 2MASS and either the DENIS or UKIDSS catalogs, 
				and histograms with respect to magnitude difference. 
				In each panel, the upper figure is for normal stars (black dots) and the bottom figure is for extracted AGNs (red dots). 
				The standard deviation for magnitude differences is shown in the upper right of each histogram. 
  \label{Corrected-Mag-Difference}}
\end{figure*}

%%%%%%%%%%%%%%%%%%%%%%%%%%%%%%%%%%%%%%%%%%%%%%%%%%
\section{Selection of near-infrared variable AGNs}\label{Selection}
\subsection{Extraction of Near-infrared Counterparts for Cataloged AGNs}
We first extracted a 2MASS counterpart for each cataloged AGN 
by cross-identifying the 2MASS PSC with the QA and SQ catalogs. 
Each 2MASS counterpart was further cross-identified with the DENIS and UKIDSS/LAS catalogs, 
and we extracted near-infrared magnitudes for different epochs. 
We only use sources having sufficient photometric quality in either the J or K bands: 
2MASS photometric quality flags superior to C [corresponding to $S/N>5$ and 
a corrected band photometry uncertainty (cmsig)$<0.21714$], 
DENIS quality flags superior to 90, 
and UKIDSS/LAS photometry having pNoise (probability that the source is noise)$\leq 0.1$, 
ErrBits$=0$, and ppErrBits$=0$. 
The ErrBits (ppErrBits) indicates processing (post-processing) error quality bit flags, 
with zero being the best quality. 
Note that all magnitudes at K$_\textnormal{\tiny S}$ (2MASS or DENIS) and K (UKIDSS) bands are denoted by $K$, 
throughout this paper. 
The number of the cataloged AGNs having near-infrared counterparts is summarized in Table \ref{Counterparts-Number}.

\begin{figure*}[thbp]
	\begin{center}
		\resizebox{80mm}{!}{\includegraphics[clip]{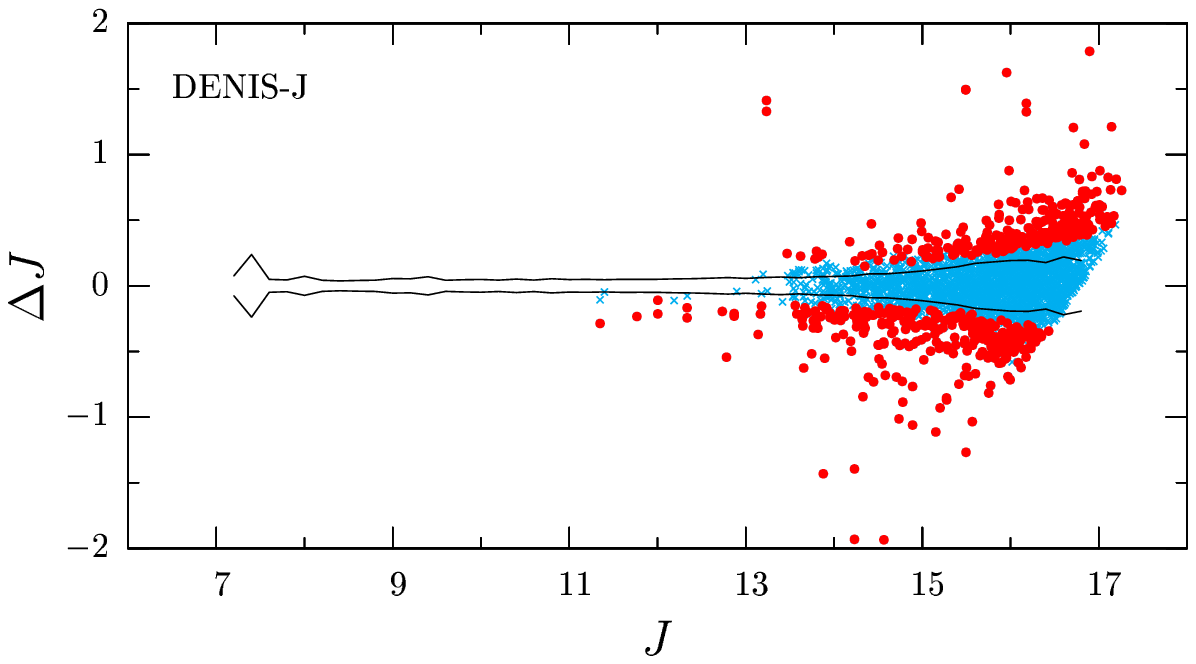}}
		\resizebox{80mm}{!}{\includegraphics[clip]{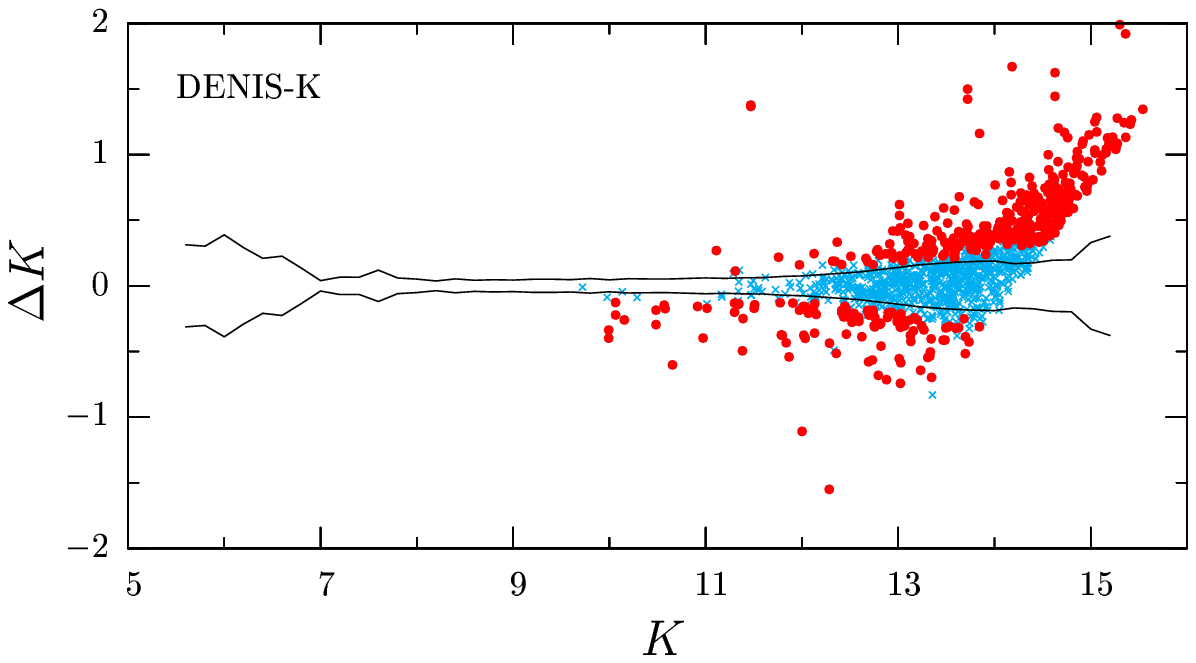}}
		\resizebox{80mm}{!}{\includegraphics[clip]{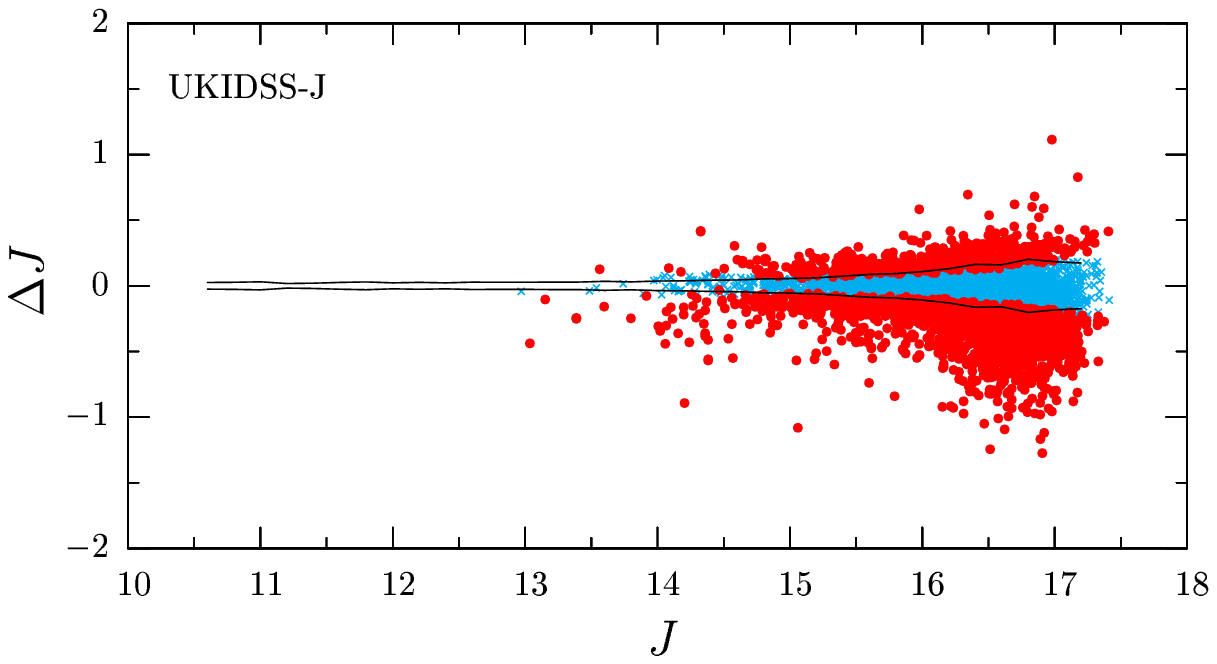}}
		\resizebox{80mm}{!}{\includegraphics[clip]{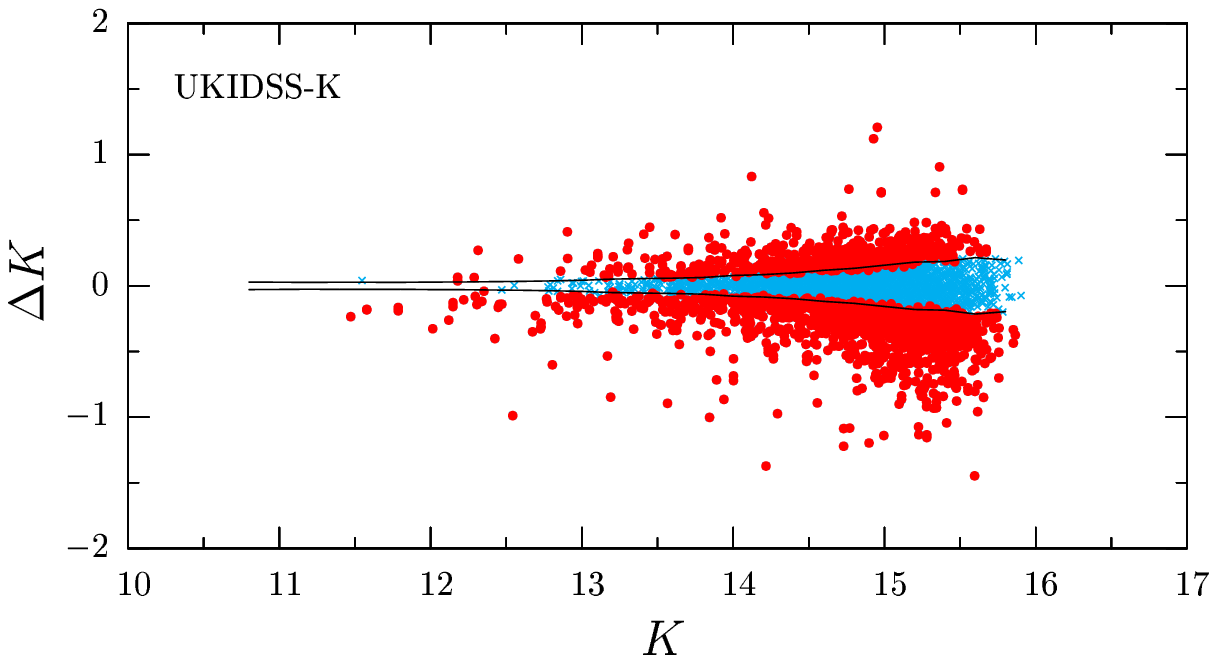}}
	\end{center}
		\caption{Magnitude versus magnitude difference for the variable (red solid circles) and non-variable (blue crosses) AGNs 
				extracted with our selection criterion. 
				The curves show $1\sigma$ confidence interval for magnitude differences of the normal stars, 
				which are derived using the sample stars in Figure \ref{Corrected-Mag-Difference}.  
  \label{Mag-Mag-Difference-AGNs}}
\end{figure*}

%%%%%%%%%%%%%%%%%%%%%%%%%%%%%%
\subsection{Zero Point of Magnitude Difference}
To examine the near-infrared variability of the cataloged AGNs, 
we use the PSF magnitudes in both the 2MASS and DENIS catalogs. 
Because there are no PSF photometry measurements in the UKIDSS/LAS catalog, 
we use average aperture magnitudes between the $2.''8$ and $5.''7$ apertures. 
\
Because we focus on the variability of AGNs by comparing the magnitudes listed in different catalogs, 
we should take into account the differences in the photometric systems between different catalogs. 
To examine a photometric difference, 
we first investigate magnitude differences of normal stars. 
We define all the magnitude differences to be 
DENIS or UKIDSS magnitudes subtracted from 2MASS magnitudes. 
We extracted a sample of 2MASS stars based on near-infrared colors of normal stars from \citet{Bessell1988-PASP}. 
The DENIS and UKIDSS magnitudes for these stars were derived by catalog cross-identification. 
We draw histograms for the magnitude differences of normal stars, 
which show Gaussian distributions with small but non-zero peak values. 
However, AGNs with near-infrared counterparts have distributions similar to those of normal stars 
but have larger peak values than normal stars. 
This might be because the AGNs are point sources but are barely more diffuse than the normal stars. 
Therefore, we corrected the magnitude differences so that the peak AGN values are coincident with 0. 
Here, we redefine the amplitude of variability for the $i$-th object as follows: 
\begin{equation}
\Delta m_i=m_{i, \textnormal{\tiny 2MASS}} - m_{i, \textnormal{\tiny DENIS or UKIDSS}} + \Delta m_{\textnormal{\tiny cor}}, 
\end{equation}
where $\Delta m_{\textnormal{\tiny cor}}$ is the correction term for the magnitude difference. 
We calculated correction terms at both the J and K bands for both the DENIS and UKIDSS catalogs (i.e., we derived four correction terms) 
and found the values to be $-0.18$ (DENIS-J), $-0.12$ (UKIDSS-J), $-0.14$ (DENIS-K), and $-0.10$ (UKIDSS-K). 

Figure \ref{Corrected-Mag-Difference} shows magnitude versus corrected magnitude difference, and histograms for each case. 
We also corrected the magnitude differences for normal stars, 
with the correction terms of $-0.02$ (DENIS-J), $-0.02$ (UKIDSS-J), $-0.03$ (DENIS-K), and $0.0$ (UKIDSS-K). 
As mentioned above, no significant difference in the histogram shape between normal stars and extracted AGNs can be seen. 
The near-infrared counterparts for the extracted AGNs are point sources 
because we extracted the counterparts from point source catalogs. 
Consequently, the similarity in the shapes is a natural result. 
Furthermore, the fact that the extracted AGNs are point sources suggests that most of the AGNs are expected to be quasars. 
\citet{Sesar2007-AJ} estimated that at least $90\%$ of quasars are variable at the 0.03 mag level. 
In all cases, the standard deviations of magnitude difference for the AGNs are larger than those for normal stars. 
This implies that most of the AGNs have a more-or-less variable nature at near-infrared wavelengths.

\begin{figure*}[thbp]
	\begin{center}
		\resizebox{75mm}{!}{\includegraphics[clip]{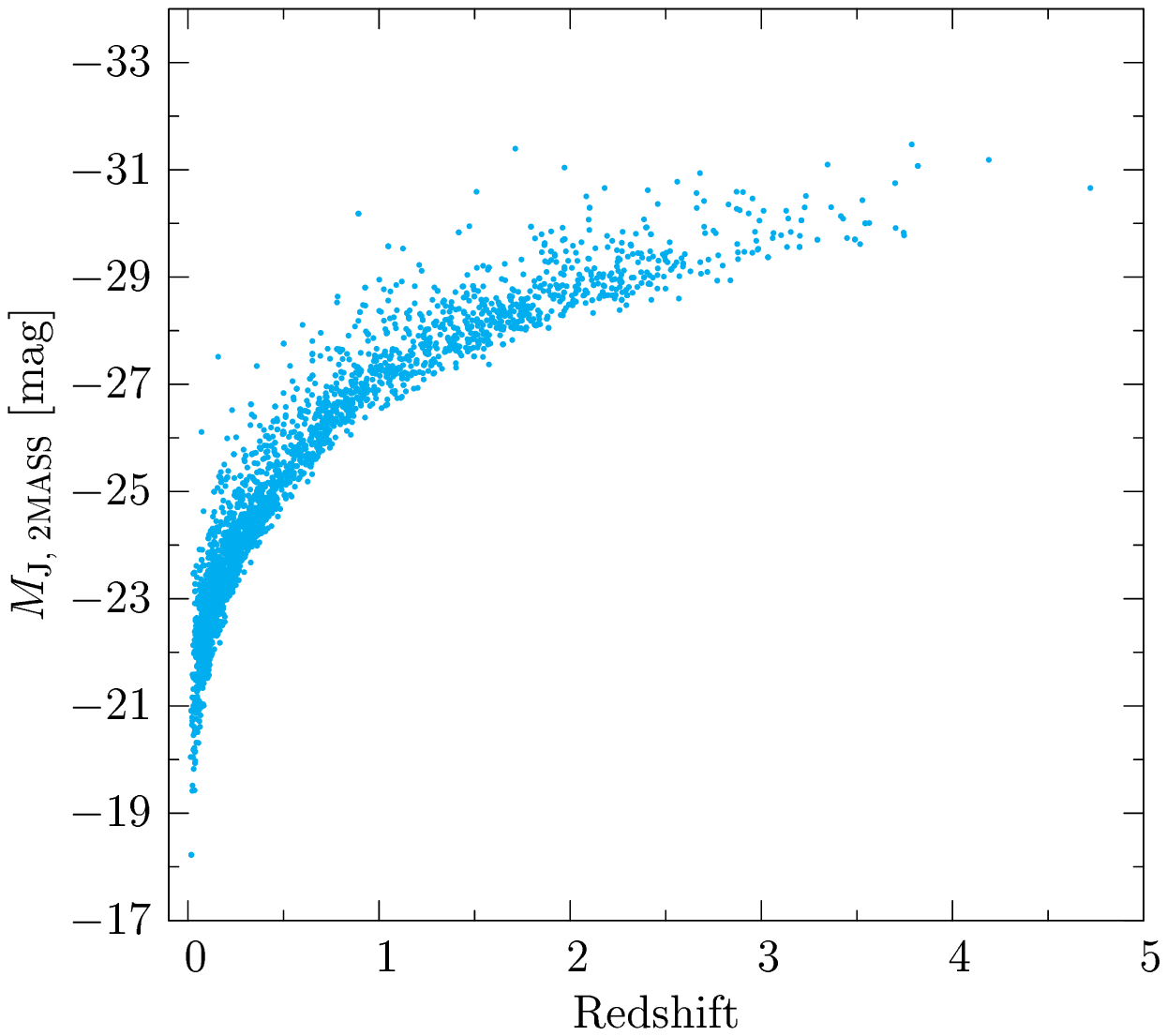}}
		\resizebox{75mm}{!}{\includegraphics[clip]{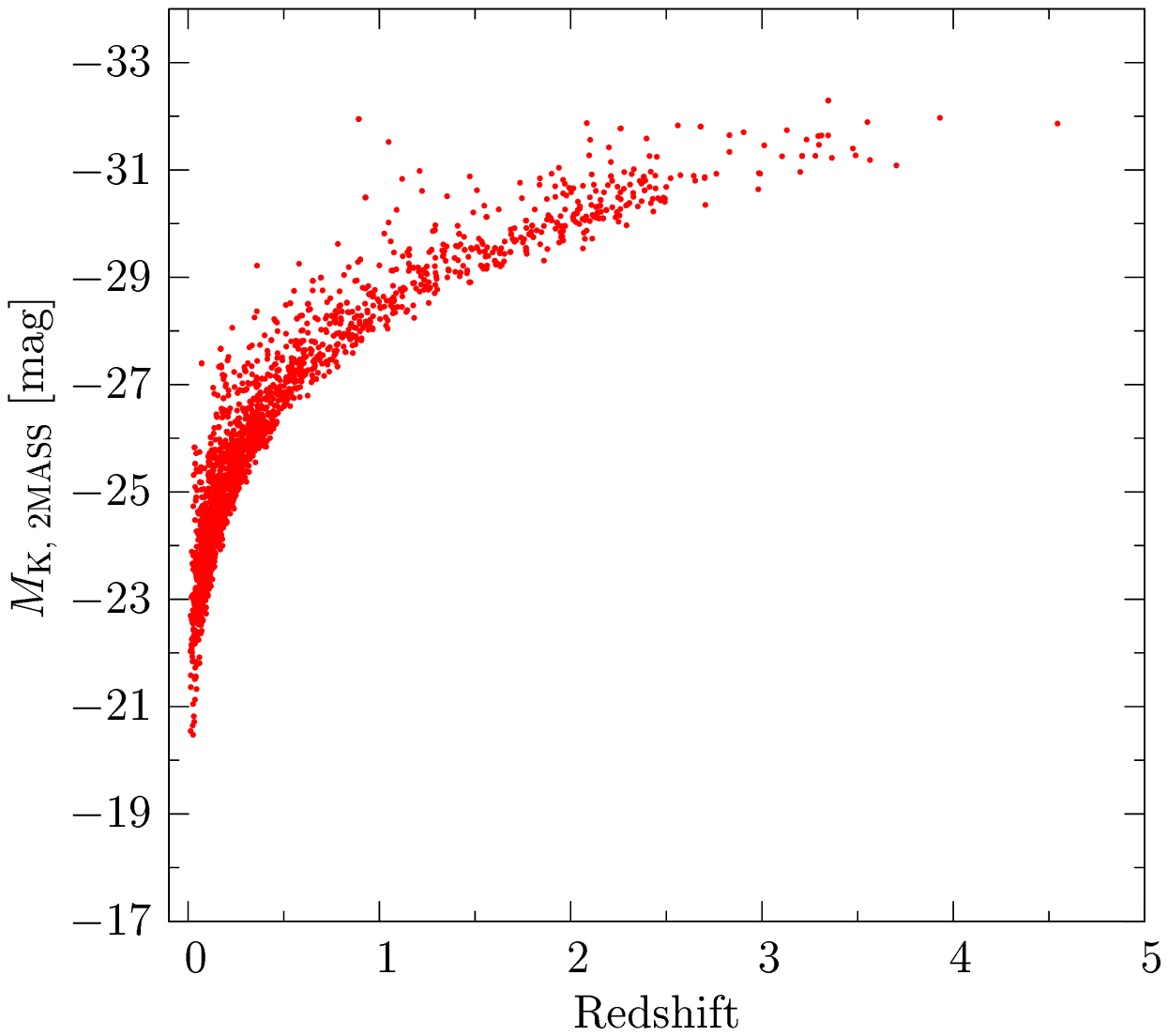}}
	\end{center}
		\caption{Redshift versus absolute magnitude. 
				The left panel is the sample in the J-band (blue), and 
				the right panel is the sample in the K-band (red). 
  \label{z-M}}
\end{figure*}

\begin{table}[thbp]
\begin{center}
\caption{Number of the selected near-infrared variable AGNs. 
		 A subset of variable AGNs is listed in both the QA and SQ catalogs. 
\label{Variable-Number}}
\begin{tabular}{lrrrr}
\hline 
 & \multicolumn{2}{c}{DENIS} & \multicolumn{2}{c}{UKIDSS} \\
 & \multicolumn{1}{c}{J} & \multicolumn{1}{c}{K} & \multicolumn{1}{c}{J} & \multicolumn{1}{c}{K} \\ \hline
QA & 554 & 426 & 1731 & 1497 \\
SQ & 105 &  55 & 1524 & 1073 \\
Total & 592 & 445 & 2316 & 1920 \\ \hline
\end{tabular}
\end{center}
\end{table}

\begin{figure*}[thbp]
	\begin{center}
		\resizebox{75mm}{!}{\includegraphics[clip]{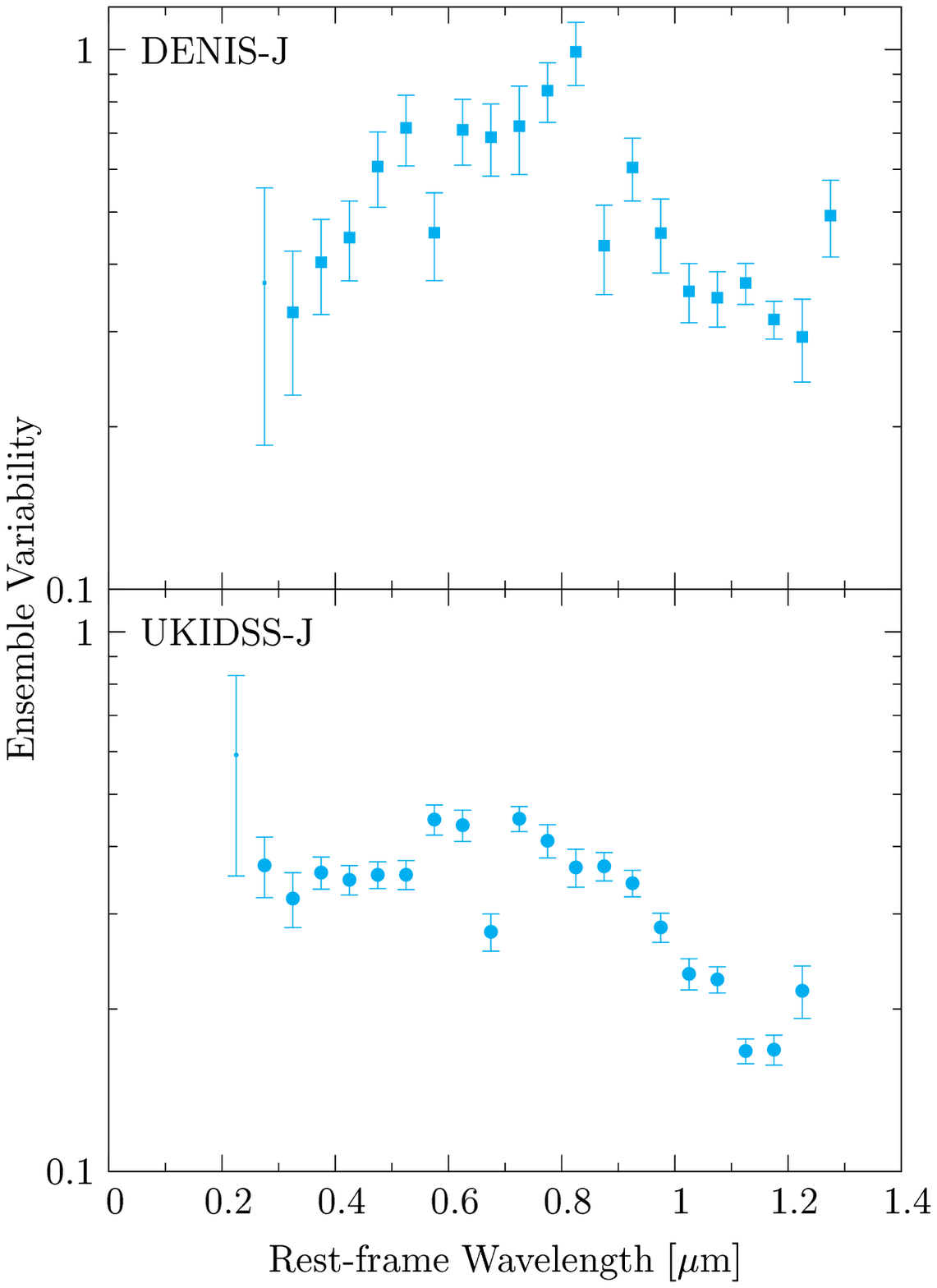}}
		\resizebox{75mm}{!}{\includegraphics[clip]{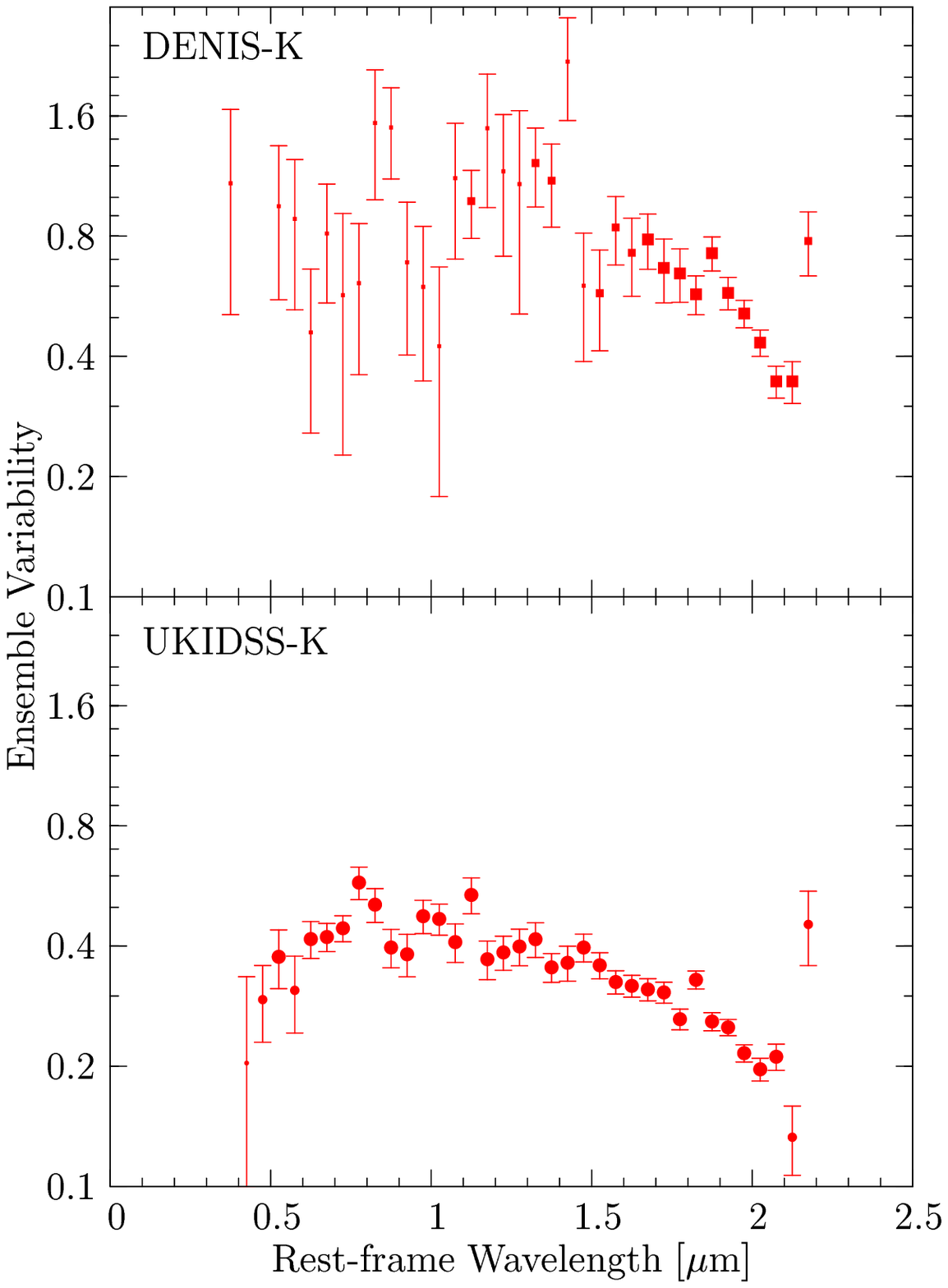}}
	\end{center}
		\caption{Variability as a function of rest-frame wavelength. 
				Colors indicate either J-selection (blue) or K-selection (red). 
				Solid squares and circles show samples from DENIS and samples from UKIDSS, respectively. 
				Three different symbol sizes represent the number of data including in each data point. 
				The largest size contains more than 9 data, the middle contains more than 4 and less than 10 data, 
				and the smallest contains less than 5 data. 
  \label{Restwavelength-Variability}}
\end{figure*}

%%%%%%%%%%%%%%%%%%%%%%%%%%%%%%
\subsection{Sample Selection}
The three near-infrared catalogs contain photometric errors. 
Taking into account these photometric errors, we selected optimal variable AGNs. 
For this paper, we regard AGNs having magnitude differences 
larger than the sum of photometric errors as near-infrared variable AGNs. 
In other words, we established the following criterion for selecting variable AGNs: 
\begin{equation}
\Delta m_i > \sigma_{i, \textnormal{\tiny 2MASS}}+\sigma_{i, \textnormal{\tiny other}}, 
\end{equation}
where $\sigma_{i, \textnormal{\tiny 2MASS}}$ is the 2MASS photometric error of the $i$-th object and 
$\sigma_{i, \textnormal{\tiny other}}$ is either the DENIS or UKIDSS photometric error of the $i$-th object. 
The selection of variable AGNs is independently performed at both the J and K bands. 
The number of near-infrared variable AGNs is summarized in Table \ref{Variable-Number}. 
Figure \ref{Mag-Mag-Difference-AGNs} shows the variable and non-variable AGNs on magnitude versus magnitude difference diagrams, 
with curves showing $1\sigma$ confidence interval for magnitude difference of the normal stars in Figure \ref{Corrected-Mag-Difference}. 
As seen in the figure, very few sources with magnitude differences smaller than $1\sigma$ level are selected as variable AGNs. 
Throughout this paper, we call the sample of variable AGNs selected in the J band (K band) as 
the J selection (K selection) or J-selected (K-selected) sample 
and call the sample derived from the DENIS (UKIDSS) catalog in the J selection (K selection) 
the DENIS (UKIDSS)-J (K) sample. 

Using the redshifts listed in the AGN catalogs, we calculated the absolute magnitudes for variable AGNs. 
Assumed cosmological constants are $H_0=71$ km s$^{-1}$ Mpc$^{-1}$, $\Omega_\textnormal{\tiny M}=0.3$, 
and $\Omega_\Lambda=0.7$. 
We applied K-corrections based on a power-law spectrum with $\alpha=0.5$. 
Redshift versus absolute magnitude relations are shown in Figure \ref{z-M}.

%%%%%%%%%%%%%%%%%%%%%%%%%%%%%%%%%%%%%%%%%%%%%%%%%%
\section{Analysis}\label{Analysis}
\subsection{Ensemble Variability}
Since we have only two sampling epochs for each AGN, 
the magnitude difference does not directly reflect the characteristic amplitude of intrinsic variability. 
In this case, the ensemble variability can be useful to examine the variability of AGN ensembles. 

Here, we define the ensemble variability as 
\begin{equation}
V=\sqrt{\frac{\sum_{i}^{N} \Delta m_i^2 - \sum_{i}^{N} \sigma_i^2}{N}}, 
\end{equation}
where $N$ is the number of variable AGNs, and 
\begin{equation}
\sigma_i^2=\sigma_{i, \textnormal{\tiny 2MASS}}^2+\sigma_{i, \textnormal{\tiny other}}^2. 
\end{equation}
The error in $V$ is represented by 
\begin{equation}
\sigma_{\tiny V}=\frac{1}{2V} \sqrt{\frac{\left( \sum_{i}^{N} \Delta m_i^2 - \sum_{i}^{N} \sigma_i^2 \right)^2}{N^3}+
\frac{\sum_{i}^{N} \left( 4 \Delta m_i^2 \sigma_i^2 - 2\sigma_i^4 \right)}{N^2}}, 
\end{equation}
where $\sigma_V$ is the measurement uncertainty. 
These are essentially the same definitions as used by \citet{Enya2002-ApJS}. 

We derive the ensemble variability for each physical parameter bin 
(i.e., rest-frame wavelength, redshift, absolute magnitude, and rest-frame time lag). 
Accordingly, the ensemble variability will be a function of a physical value. 
Below, we examine the correlations between the ensemble variability and the physical parameters.

\begin{figure*}[thbp]	
	\begin{center}
		\resizebox{75mm}{!}{\includegraphics[clip]{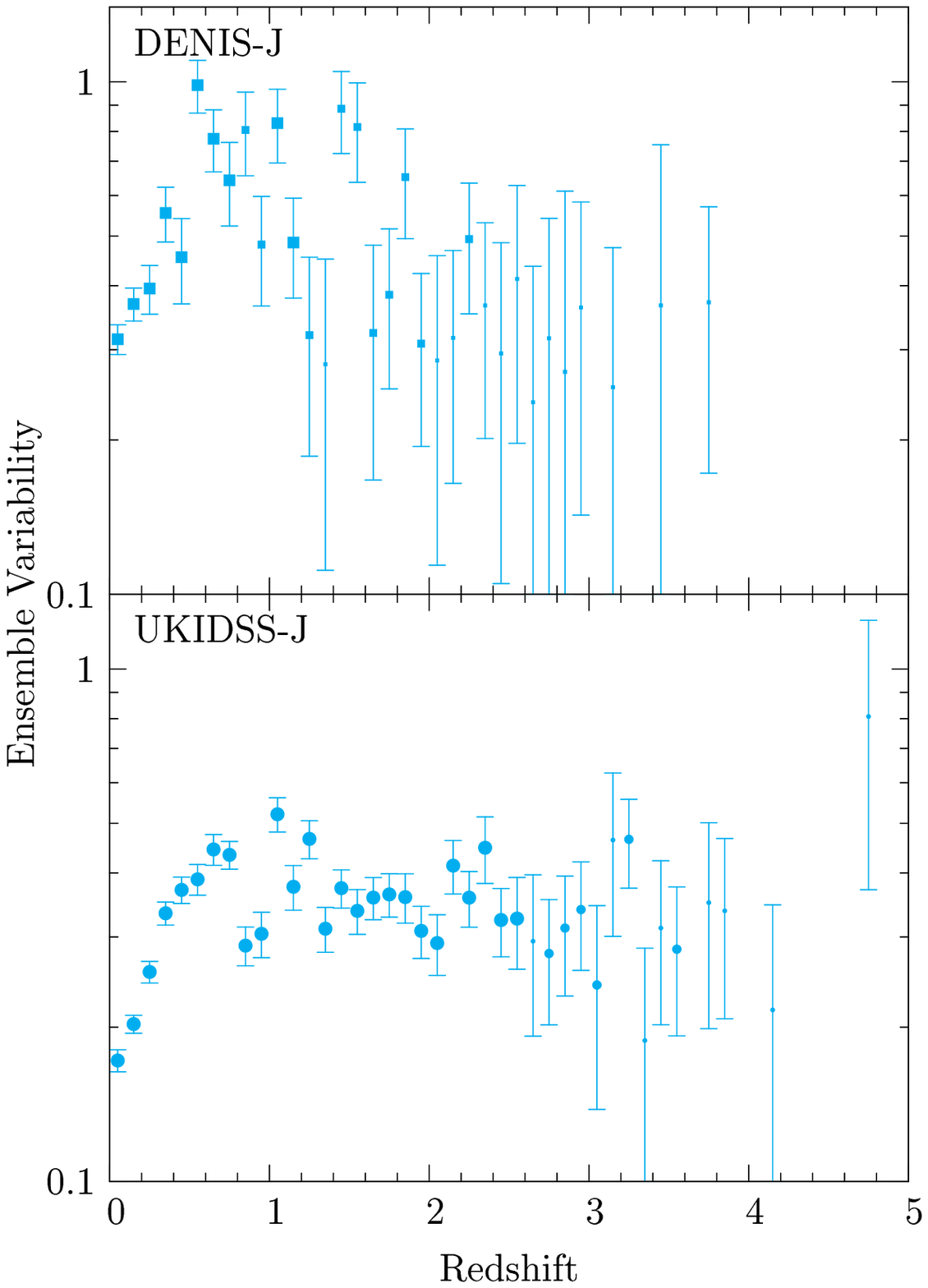}}
		\resizebox{75mm}{!}{\includegraphics[clip]{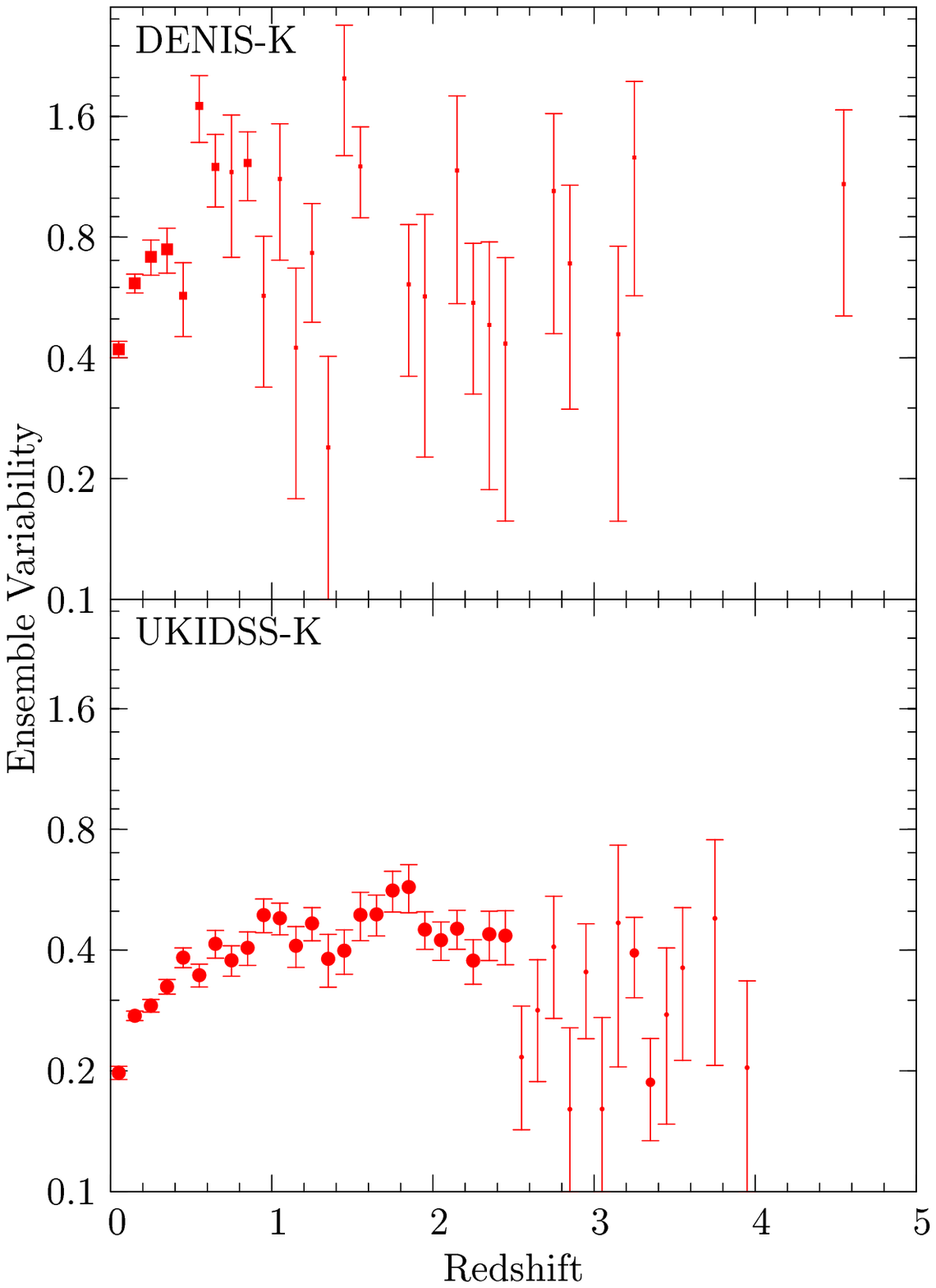}}
	\end{center}
		\caption{Variability as a function of redshift. The symbols are the same as in Figure \ref{Restwavelength-Variability}. 
  \label{Redshift-Variability}}
\end{figure*}

%%%%%%%%%%%%%%%%%%%%%%%%%%%%%%
\subsection{Rest-frame Wavelength}\label{Rest-frame-Wavelength}
Quasar variability is known to be related to the rest-frame wavelength. 
In the optical, many previous studies have demonstrated that the variability decreases with increasing rest-frame wavelength 
\citep{Giallongo1991-ApJ,Cristiani1996-AA,Fernandes1996-MNRAS,Trevese2002-ApJ,VandenBerk2004-ApJ}. 
In addition, anticorrelation has been established to span at least from the ultraviolet to near-infrared 
\citep{Cutri1985-ApJ,Paltani1994-AA,Clemente1996-ApJ,Cristiani1997-AA,Trevese2002-ApJ,Meusinger2011-AA}. 

Figure \ref{Restwavelength-Variability} shows variability as a function of rest-frame wavelength. 
For all the samples, a clear trend in the ensemble variability is to decrease with increasing rest-frame wavelength in the near-infrared 
(i.e., $\lambda_{\textnormal{\tiny rest}} \gtrsim 0.8$). 
Although the data points for the DENIS-K sample are unreliable 
in the range of $\lambda_{\textnormal{\tiny rest}} \lesssim 1.5$ because of low statistics, 
an anticorrelation over at least $\lambda_{\textnormal{\tiny rest}} \sim 1.5$ is clearly evident. 
This general trend suggests that the anticorrelation between variability and rest-wavelength is established 
not only in the optical but also in the near-infrared. 

Conversely, in the optical (i.e., $\lambda_{\textnormal{\tiny rest}} \lesssim 0.8$), 
variability appears to increase with increasing rest-frame wavelength. 
A turnover can be seen around $\lambda_{\textnormal{\tiny rest}}=0.8$ in the DENIS-J sample. 
This is significant because the Spearman rank-correlation coefficient is 0.891 ($-0.576$) for the positive (negative) slope 
in $\lambda_{\textnormal{\tiny rest}}<0.8$ ($\lambda_{\textnormal{\tiny rest}}\geq0.8$), with the significant at 99\% (91\%) confidence level. 
Only the UKIDSS-J sample shows a moderately flat curve. 
These trends in the optical are inconsistent with previous studies. 
A possible cause of these conflicting results is biased sampling of high-redshift AGNs. 
AGNs with higher redshifts become more difficult to detect because their apparent magnitudes should be more faint. 
Even if an AGN could be detected at its brighter phase, 
the magnitude is expected to be near the limiting magnitudes of the near-infrared catalogs. 
In most cases, the faint-phase magnitude of the AGN is expected to be fainter than the limiting magnitudes. 
Because we selected variable AGNs by comparing only double-epoch magnitudes, 
we will not be able to, in such a case, extract the AGN as a variable source. 
In particular, AGNs having larger variability tend to fall into this situation and 
therefore cannot be selected as variable AGNs. 
Accordingly, we could not extract variable AGNs having large variability amplitudes even if such AGNs exist, and 
most variable AGNs we could extract should have relatively small variability amplitudes. 
Because AGNs with higher redshifts are detected at shorter wavelengths in the rest-frame for a fixed passband in the observer-frame, 
this biased sampling would especially affect the correlation at shorter optical wavelengths. 

Indeed, the steeper slope for the DENIS-J sample than that for the UKIDSS-J sample 
can be interpreted as the difference between limiting magnitudes. 
The DENIS cannot detect fainter AGNs in the same way as the UKIDSS, 
because the limiting magnitudes of the DENIS are brighter than that of the UKIDSS. 
As a result, at shorter rest-frame wavelengths, 
the number of high-redshift AGNs showing large variability in the DENIS-J sample 
becomes smaller than that in the UKIDSS-J sample due to the above-mentioned biased sampling. 
This might cause the steeper slope in the DENIS-J sample.

\begin{figure*}[thbp]
	\begin{center}
		\resizebox{40mm}{!}{\includegraphics[clip]{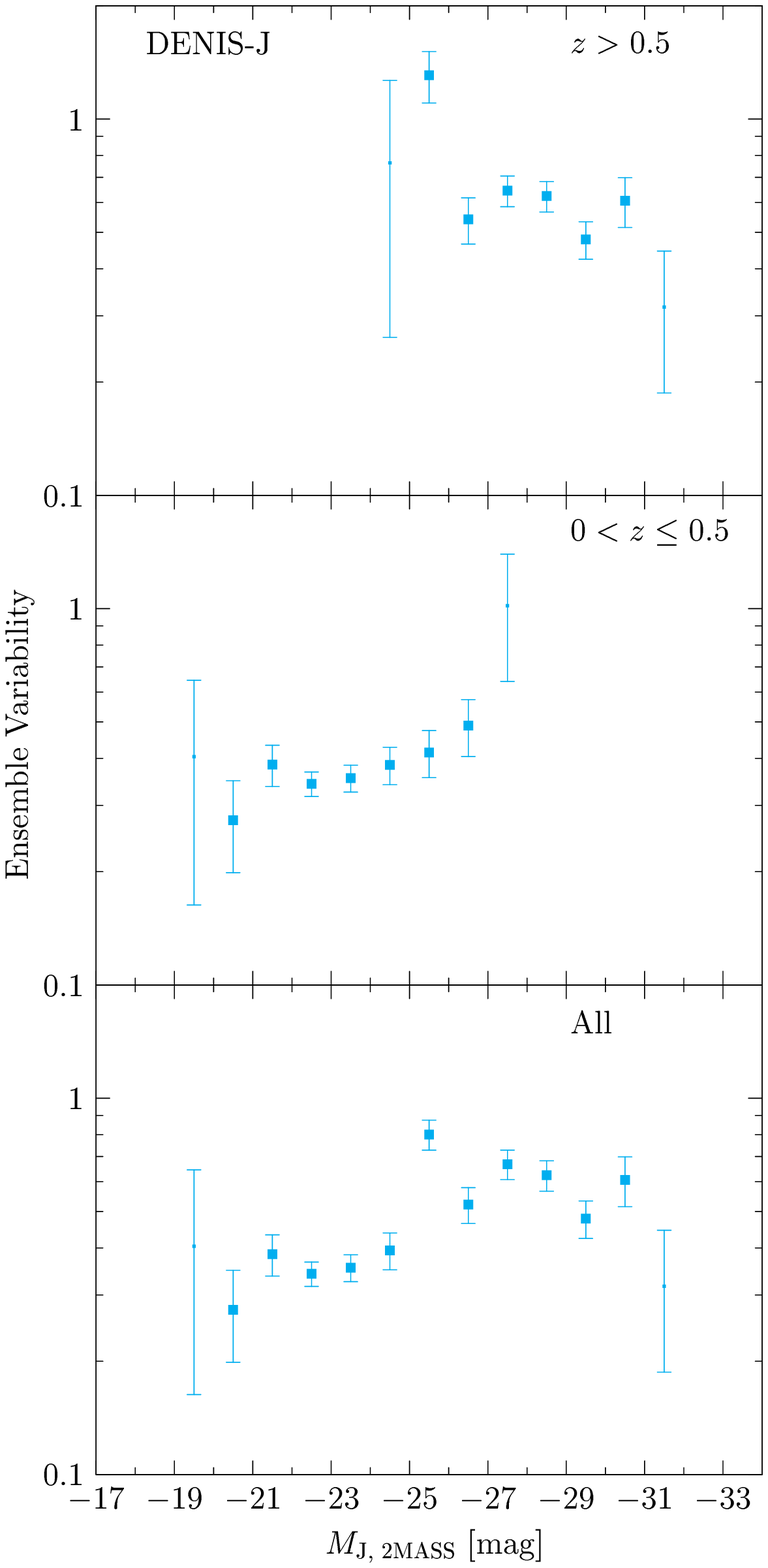}}
		\resizebox{40mm}{!}{\includegraphics[clip]{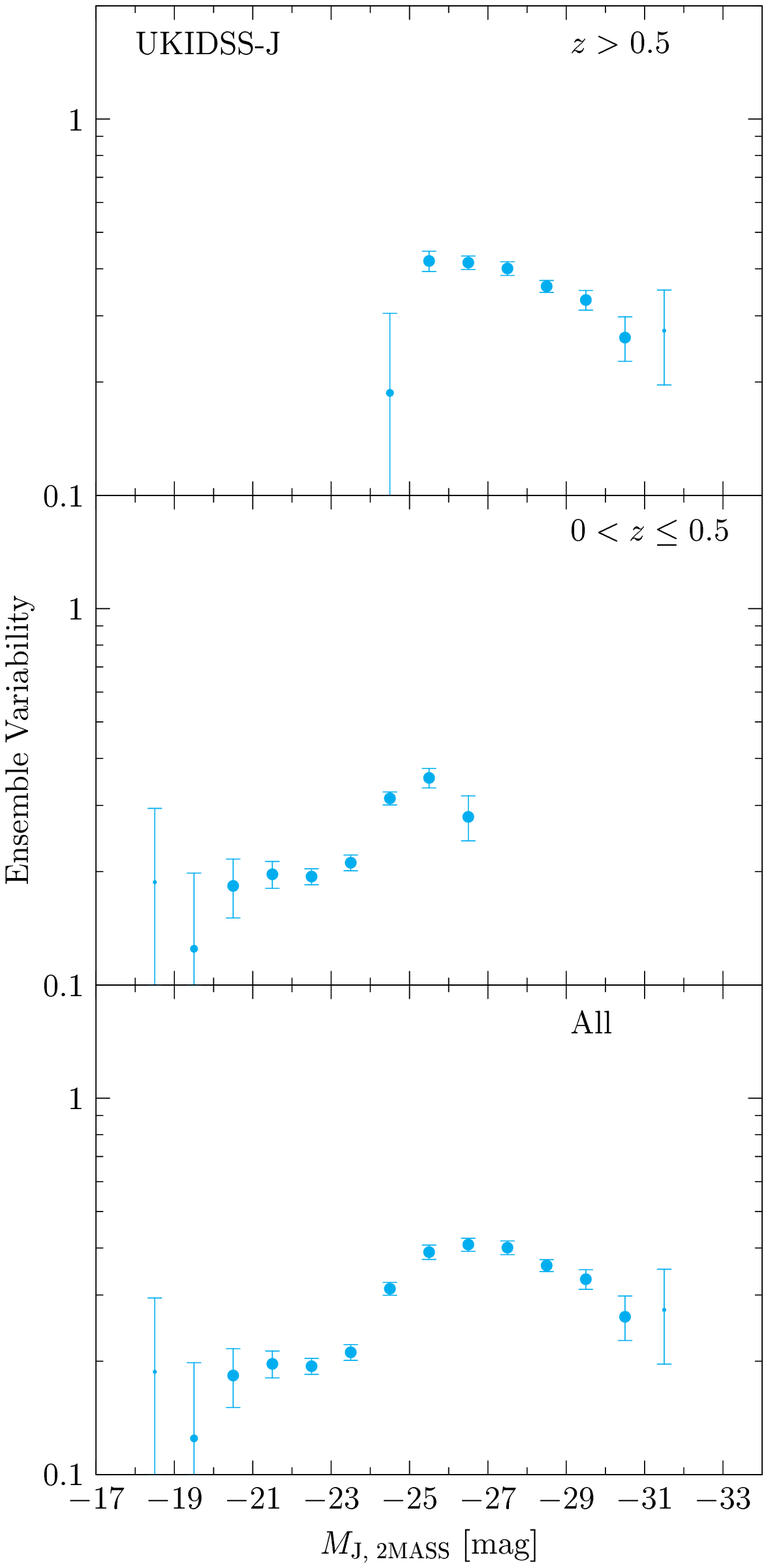}}
		\resizebox{40mm}{!}{\includegraphics[clip]{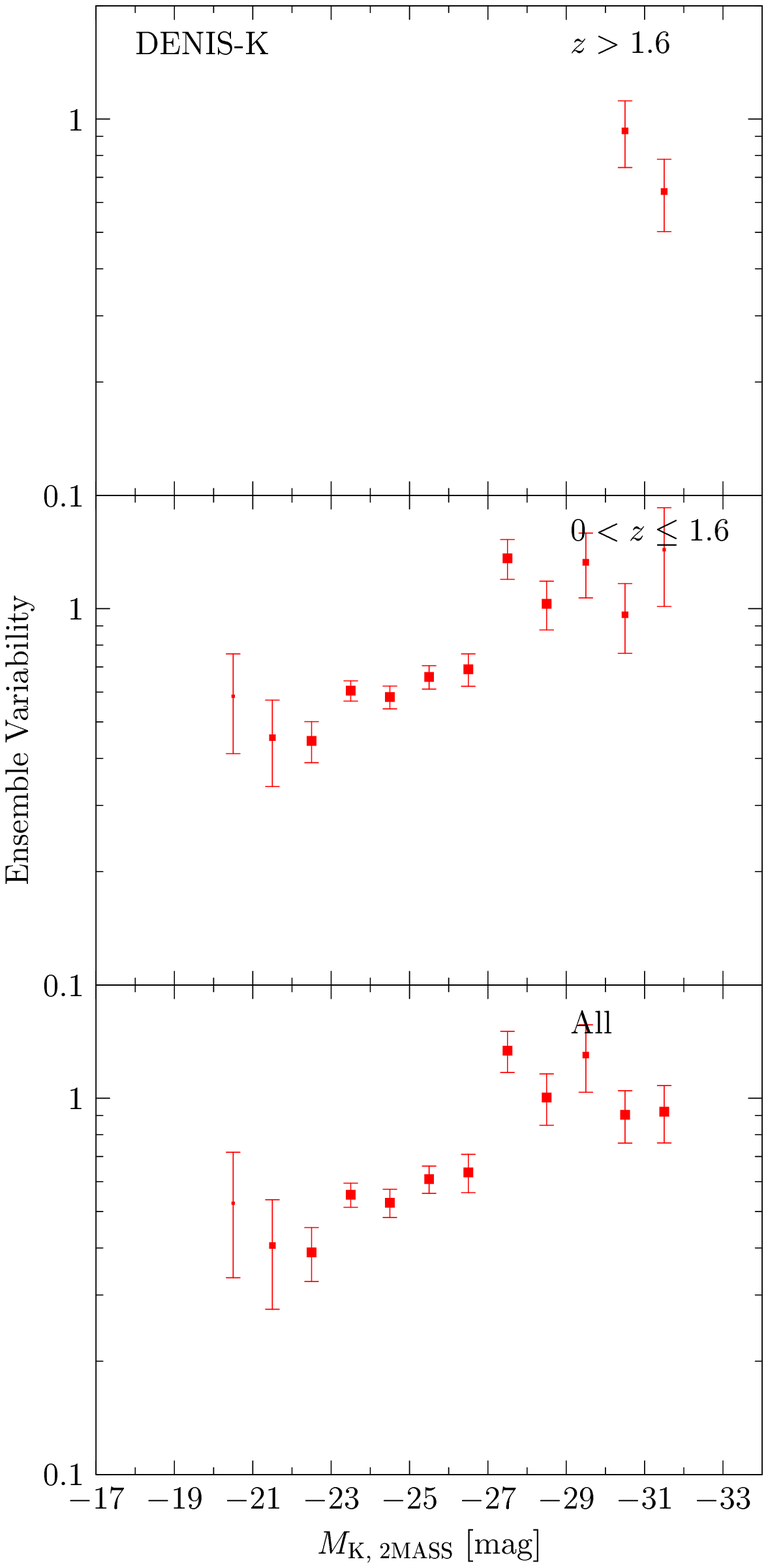}}
		\resizebox{40mm}{!}{\includegraphics[clip]{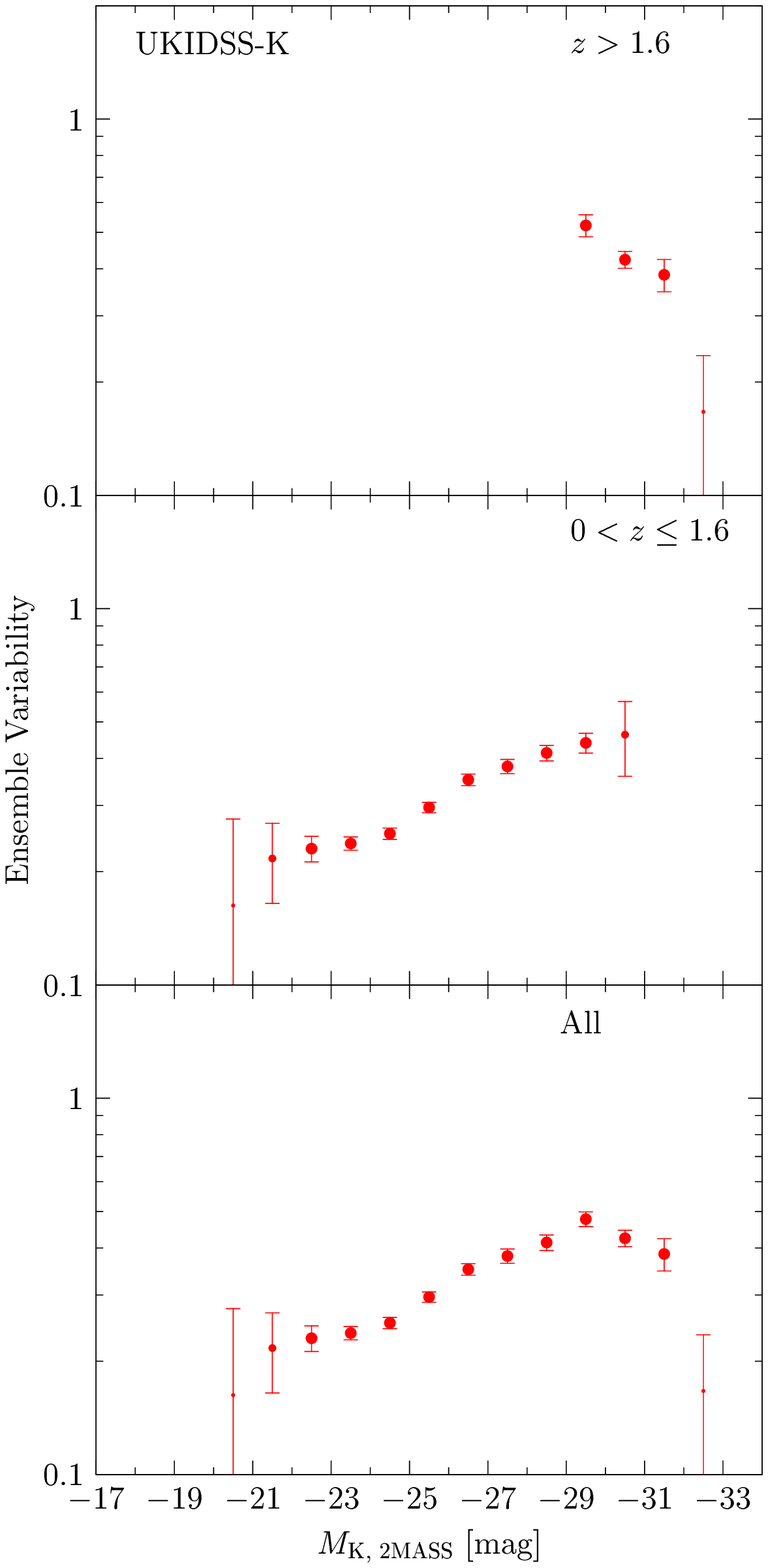}}
	\end{center}
		\caption{Variability as a function of near-infrared absolute magnitude. 
				The symbols are the same as for Figure \ref{Restwavelength-Variability}. 
				The J-selected (K-selected) samples are separated at $z=0.5$ ($z=1.6$). 
  \label{M-Variability}}
\end{figure*}

%%%%%%%%%%%%%%%%%%%%%%%%%%%%%%
\subsection{Redshift}\label{Redshift}
Correlations between redshift and variability are shown in Figure \ref{Redshift-Variability}. 
In the UKIDSS-J sample, variability increases with increasing redshift below $z\sim 0.8$. 
The DENIS-J sample also shows an increase similar to that of the UKIDSS-J sample. 
However, the slopes appear to change significantly around $z = 0.6$--$0.8$. 
Above $z\sim 0.8$, there seems to be an anticorrelation in both samples, 
although many data points are unreliable because of the small number of objects contributing to the data 
(this is especially true for the DENIS-J sample). 
The J-band wavelength in the observer-frame corresponds to a wavelength of 0.70--0.78 $\mu$m in the rest-frame at $z=0.6$--$0.8$. 
This is a near boundary wavelength between the optical and the near-infrared regions. 

The K-selected samples also show a property similar to the J-selected samples. 
The UKIDSS-K sample shows a positive correlation below $z\sim 2$. 
In the rest-frame at this redshift, 
the K-band wavelength in the observer-frame also corresponds to a near boundary wavelength 
between the optical and the near-infrared. 
Beyond $z\sim 2$, there seems to be a negative correlation. 
No trend can be found in the DENIS-K sample because most data points are scattered 
because of the statistically small number of data at each point. 

\citet{Enya2002-ApJS} reported that the sample of radio-loud AGNs shows a positive correlation 
between near-infrared variability and redshift, 
whereas the sample of radio-quiet AGNs shows a negative correlation. 
According to their correlations, 
our sample AGNs exhibit a property similar to their radio-loud AGNs. 

Note that the correlations of variability with redshift can be interpreted 
as an anticorrelation between wavelength and variability \citep{Cristiani1996-AA,Fernandes1996-MNRAS,Clemente1996-ApJ,Cristiani1997-AA}. 
In other words, for a fixed passband in the observer-frame, AGNs with larger redshifts are detected at shorter wavelengths, 
which have systematically large fluctuations. 
In this case, the correlation of variability with redshift does not reflect evolutionary effects. 
Therefore, the variations in the slope near the boundary wavelength between the optical and near-infrared 
should be intrinsic to the variations in the slope mentioned in Section \ref{Rest-frame-Wavelength}.

\begin{figure*}[thbp]	
	\begin{center}
		\resizebox{40mm}{!}{\includegraphics[clip]{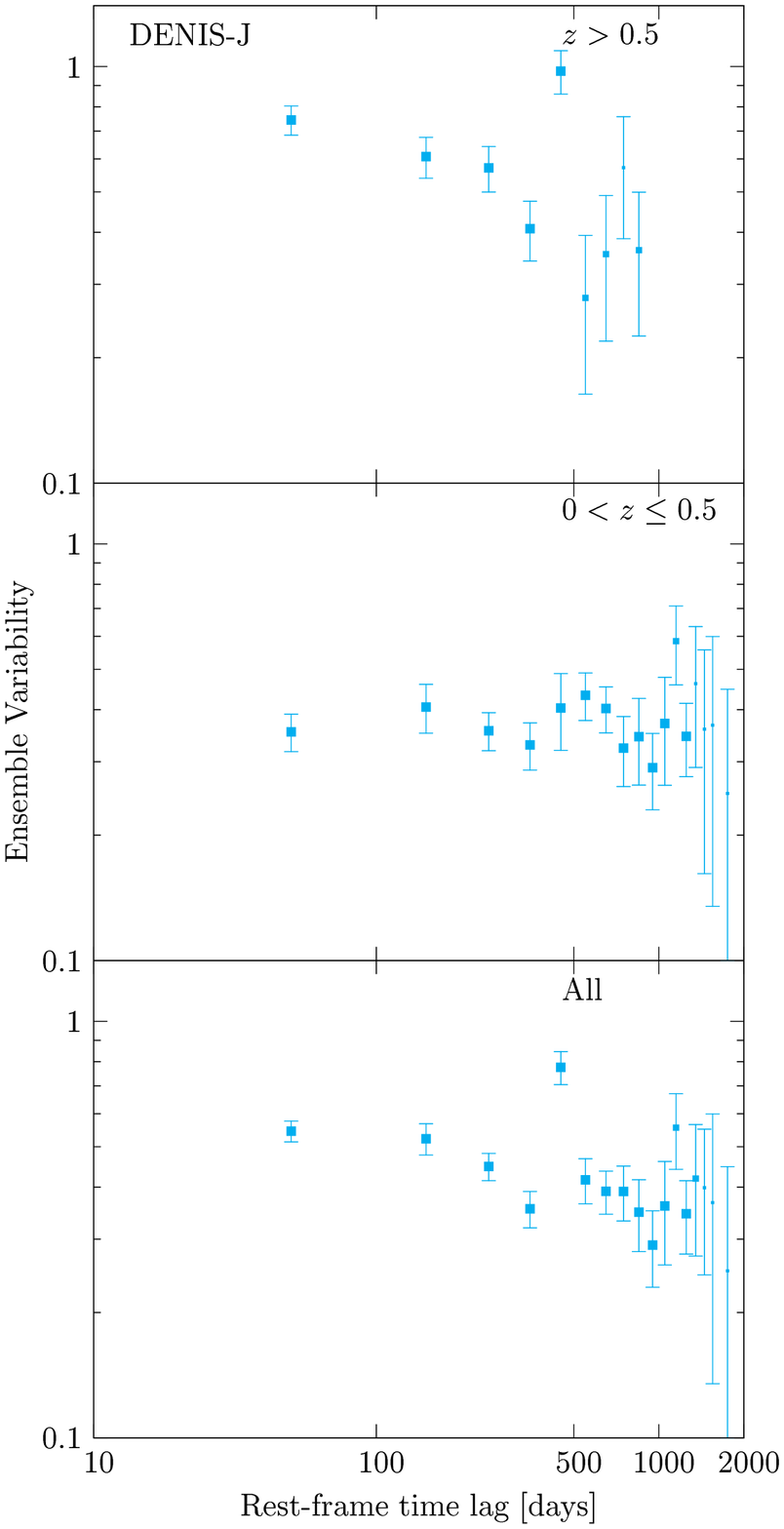}}
		\resizebox{40mm}{!}{\includegraphics[clip]{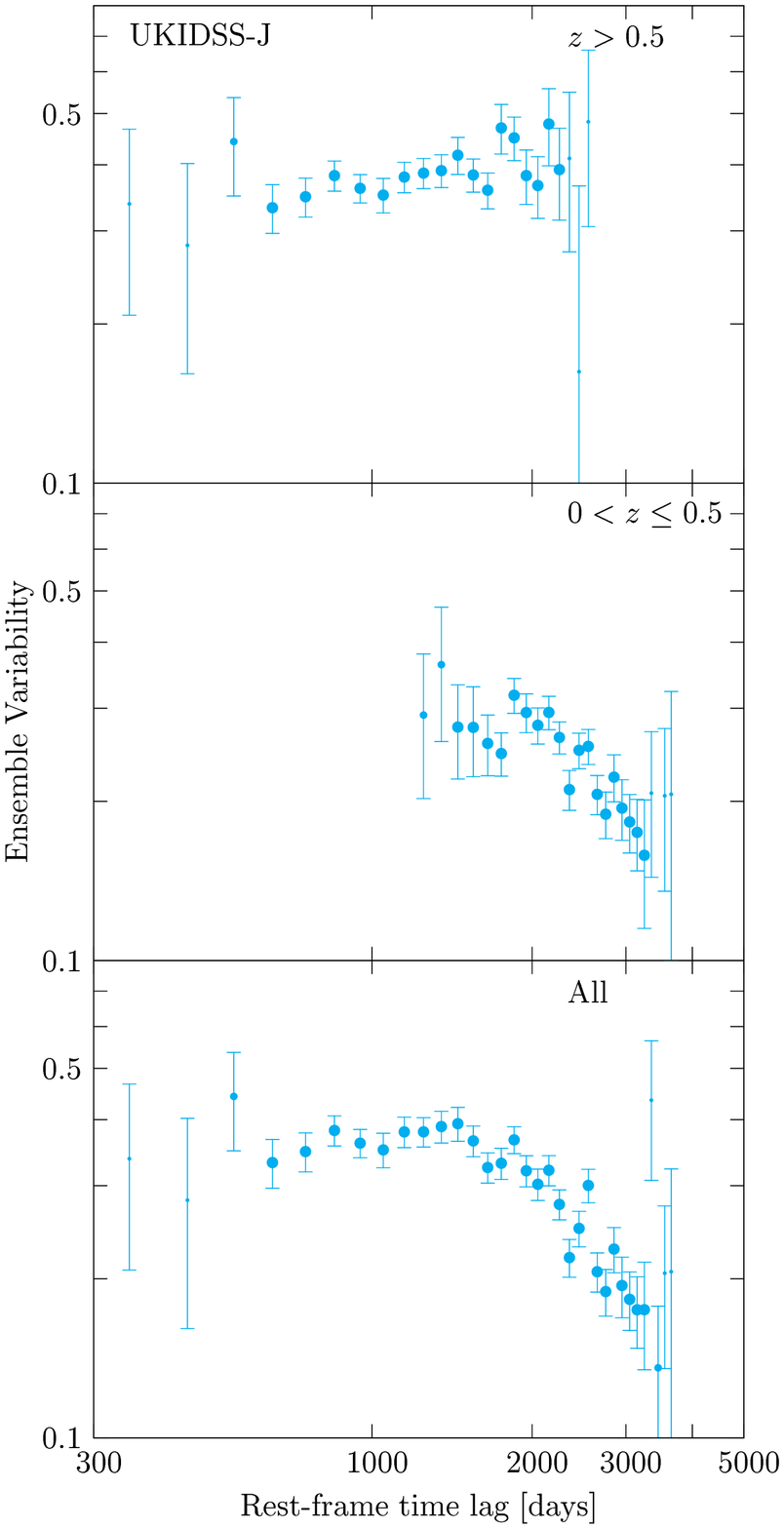}}
		\resizebox{40mm}{!}{\includegraphics[clip]{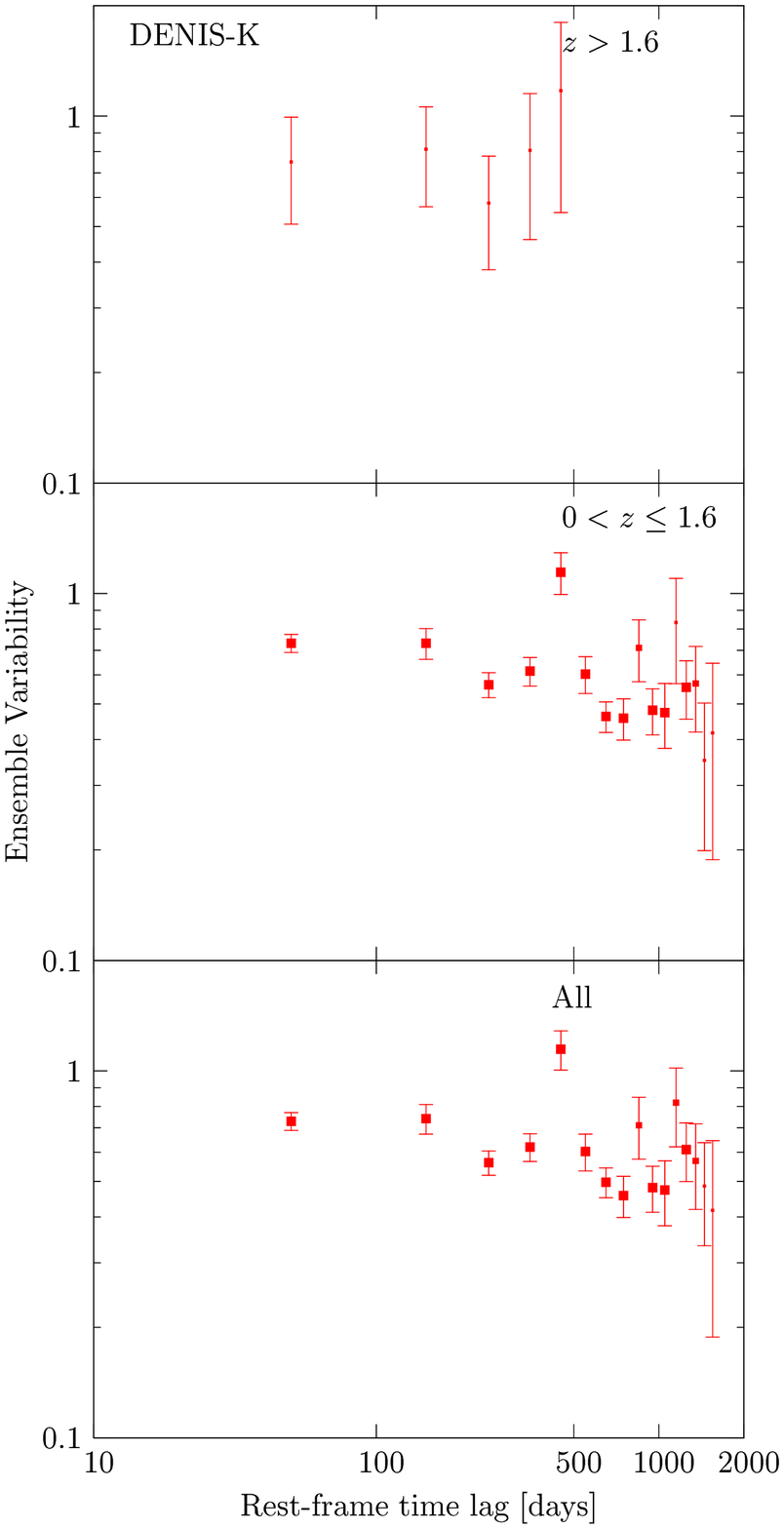}}
		\resizebox{40mm}{!}{\includegraphics[clip]{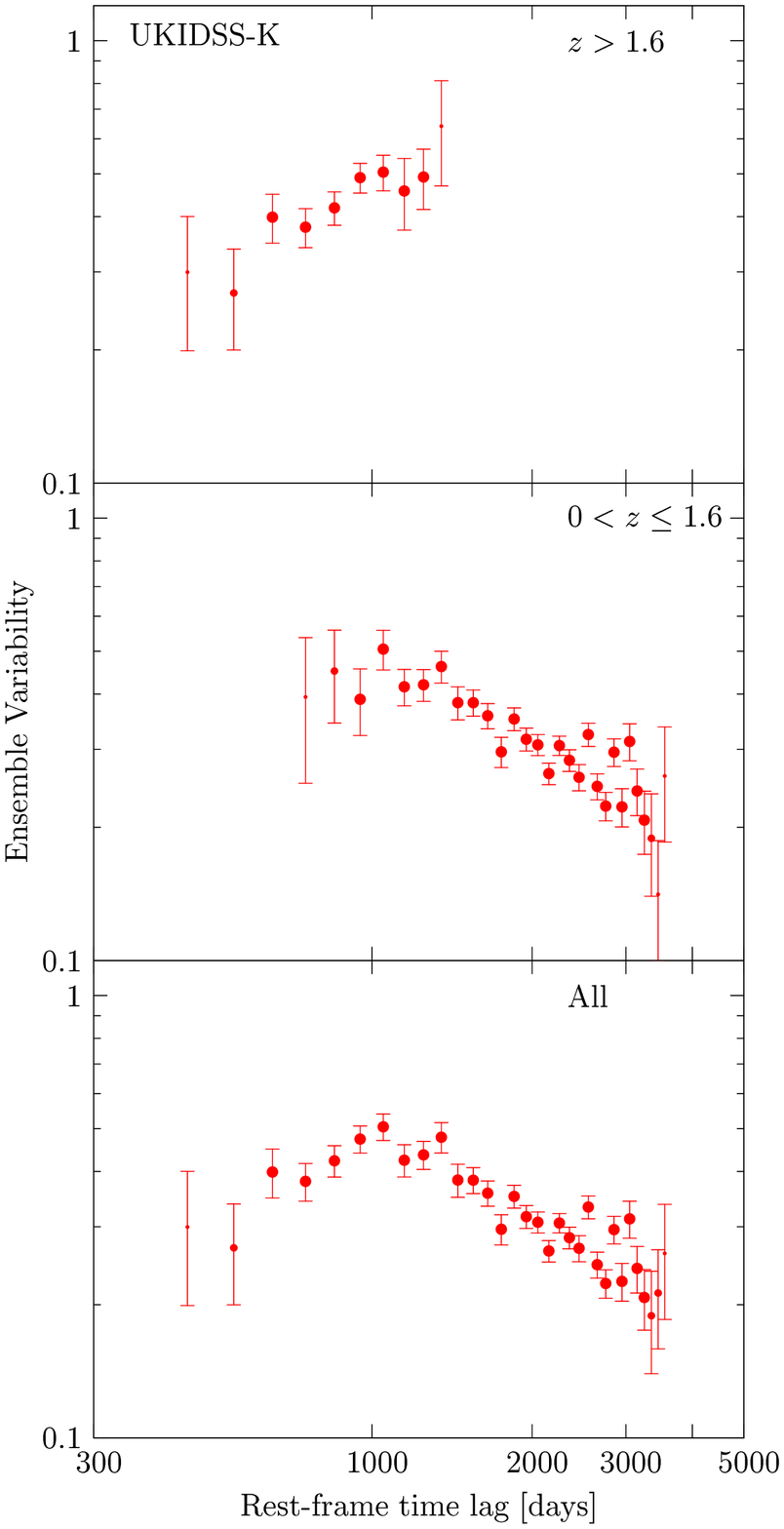}}
	\end{center}
		\caption{Structure functions for each sample. The symbols are same as Figure \ref{Restwavelength-Variability}. 
				The J-selected (K-selected) samples are separated at $z=0.5$ ($z=1.6$). 
  \label{SF}}
\end{figure*}

%%%%%%%%%%%%%%%%%%%%%%%%%%%%%%
\subsection{Absolute Magnitude}\label{Absolute-Magnitude}
As shown above, the intrinsic variability properties in the optical appear to be different from those in the near-infrared. 
Consequently, in what follows, we divide our J-selected (K-selected) samples into two subsets according to 
whether an object's redshift is $z\leq0.5$ or $z>0.5$ ($z\leq1.6$ or $z>1.6$). 
These boundary redshifts correspond to a rest-frame wavelength of about $0.8$ $\mu$m, 
which is a near boundary wavelength between the optical and near-infrared. 
When the redshift of an object is larger than the boundary redshift 
(i.e., $z=0.5$ or $z=1.6$ for J- or K-selected samples, respectively), 
the variability of the object represents the intrinsic optical variability. 

Figure \ref{M-Variability} shows correlations between absolute magnitude (luminosity) and ensemble variability. 
For ``all'' of the J-selected samples (see bottom-row panels in Figure \ref{M-Variability}), 
a broad peak appears roughly at $M_{\textnormal{\tiny J}} \sim -27$. 
In other words, the ensemble variability increases with decreasing absolute magnitude (i.e., negative correlation) 
for $M_{\textnormal{\tiny J}} \gtrsim -27$, 
whereas there is positive correlation for $M_{\textnormal{\tiny J}} \lesssim -27$. 
As can be seen from the middle- and top-row panels in Figure \ref{M-Variability}, 
the slope variations are caused by a difference in variability between the optical and the near-infrared. 
The K-selected samples also show similar trends; 
that is, a negative correlation appears in the near-infrared (i.e., $0<z\leq1.6$) samples 
whereas a positive correlation appears in the optical (i.e., $z>1.6$) samples. 

Various studies have confirmed that the optical variability of quasars is anticorrelated with luminosity 
\citep{Uomoto1976-AJ,Hook1994-MNRAS,Cristiani1996-AA,Giveon1999-MNRAS,VandenBerk2004-ApJ}, 
which indicates that more-luminous quasars exhibit smaller variability. 
The optical variability of our samples is consistent with these analyses. 

In Poissonian models, 
the following relationship is expected between fractional optical variability $\Delta L/L$ and mean luminosity $L$: 
\begin{equation}\label{V-L}
\frac{\Delta L}{L} \propto L^{-\beta}, 
\end{equation}
where $\beta=\frac{1}{2}$ \citep{Fernandes2000-ApJ}. 
Equation (\ref{V-L}) is equivalent to 
\begin{equation}
\log (V) = \frac{\beta}{2.5} M + K, 
\end{equation}
where $M$ is the absolute magnitude and $K$ is a constant. 
Using this relationship, we measure, by weighted least square fits, 
$\beta=$ 0.096 (DENIS-J), 0.080 (UKIDSS-J), 0.396 (DENIS-K), and 0.167 (UKIDSS-K). 
Only the DENIS-K sample has a steeper slope, which is considerably steeper than the other three samples. 
Despite this result, we expect the DENIS-K slope to be less reliable than the others because it is calculated from only two data points. 
The slopes of the other three samples are less than 0.2, 
which is inconsistent with the Poissonian prediction. 
Additionally, the three slopes are also less than such as $\beta=0.246 \pm 0.005$ \citep{VandenBerk2004-ApJ} and 
$\beta=0.205 \pm 0.002$ \citep{Bauer2009-ApJ}. 
We consider it highly probable that the more-level slopes are due to the same cause as mentioned in Section \ref{Rest-frame-Wavelength}; 
namely, to a biased sampling of large-redshift AGNs. 

In contrast with the optical correlations, 
the near-infrared variability is negatively correlated with absolute magnitude for the whole sample. 
The slope of each sample is $-0.062$ (DENIS-J), $-0.156$ (UKIDSS-J), $-0.093$ (DENIS-K), and $-0.120$ (UKIDSS-K). 
In addition, these negative correlations are also qualitatively consistent with the trend of radio-loud AGNs reported by \citet{Enya2002-ApJS}. 

The near-infrared correlations can be explained by the calculations of \citet{Kawaguchi2011-pre}, 
which predict that thinner tori have larger near-infrared variability amplitudes (see Figure 10 in their paper). 
Assuming more luminous AGNs have thinner tori [i.e., the receding torus model \citep[e.g.,][]{Lawrence1991-MNRAS}], 
their model implies that more-luminous AGNs have larger near-infrared variability.

\begin{figure*}[thbp]
	\begin{center}
		\resizebox{75mm}{!}{\includegraphics[clip]{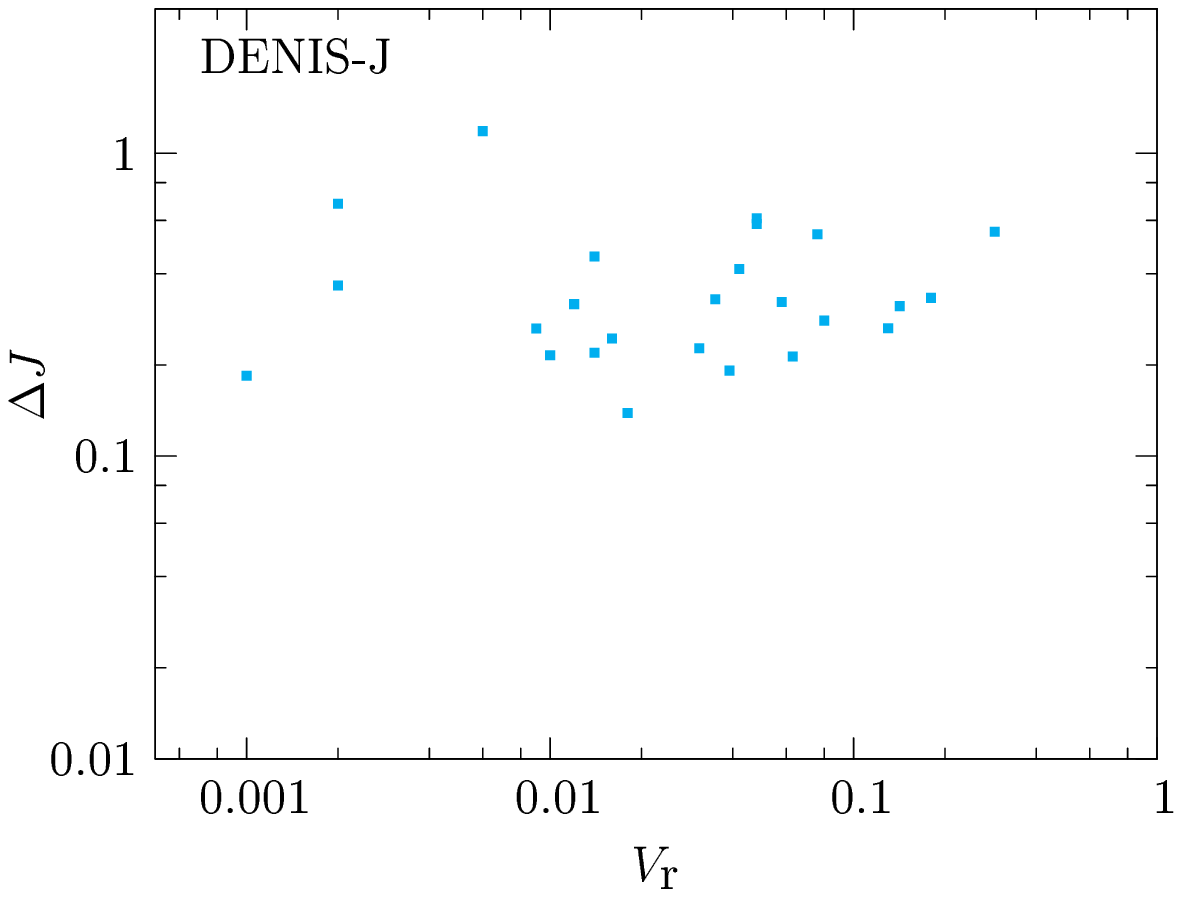}}
		\resizebox{75mm}{!}{\includegraphics[clip]{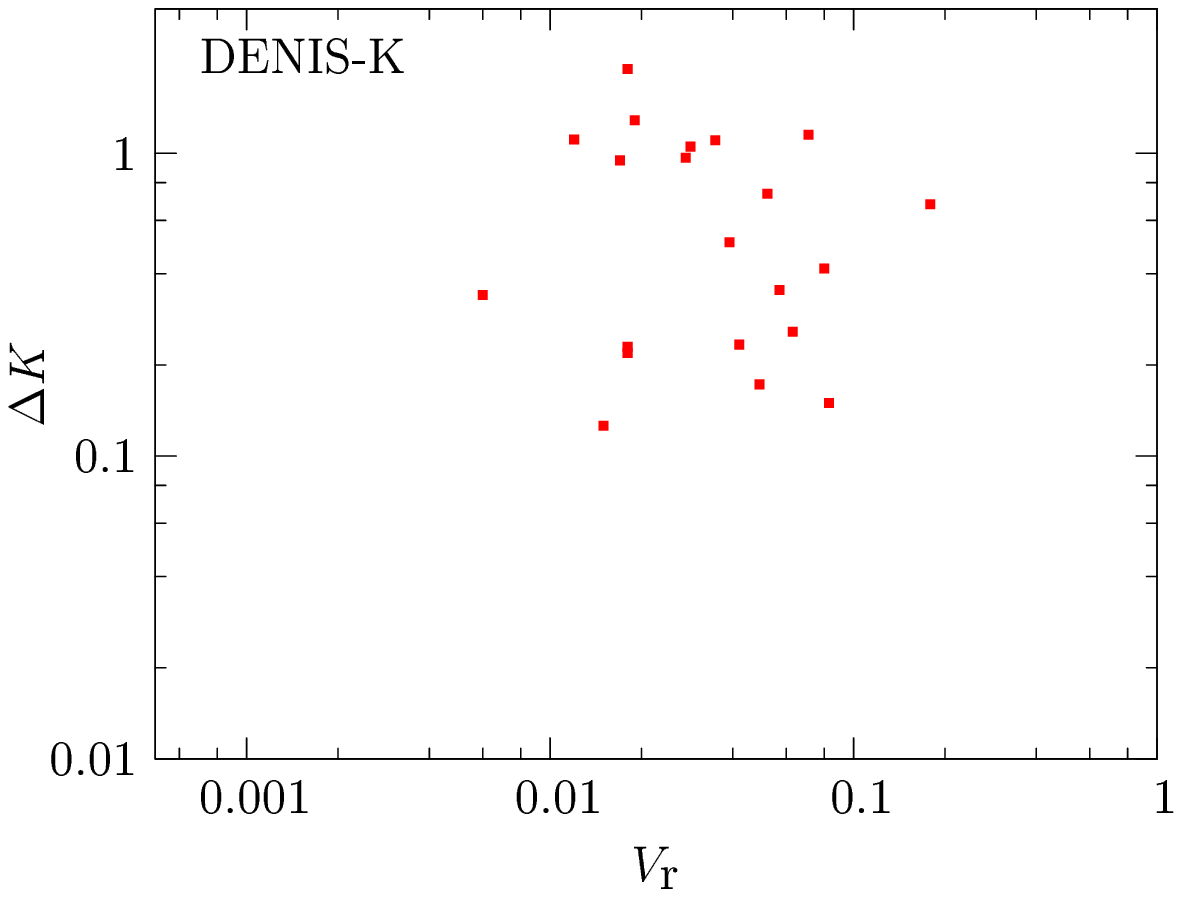}}
		\resizebox{75mm}{!}{\includegraphics[clip]{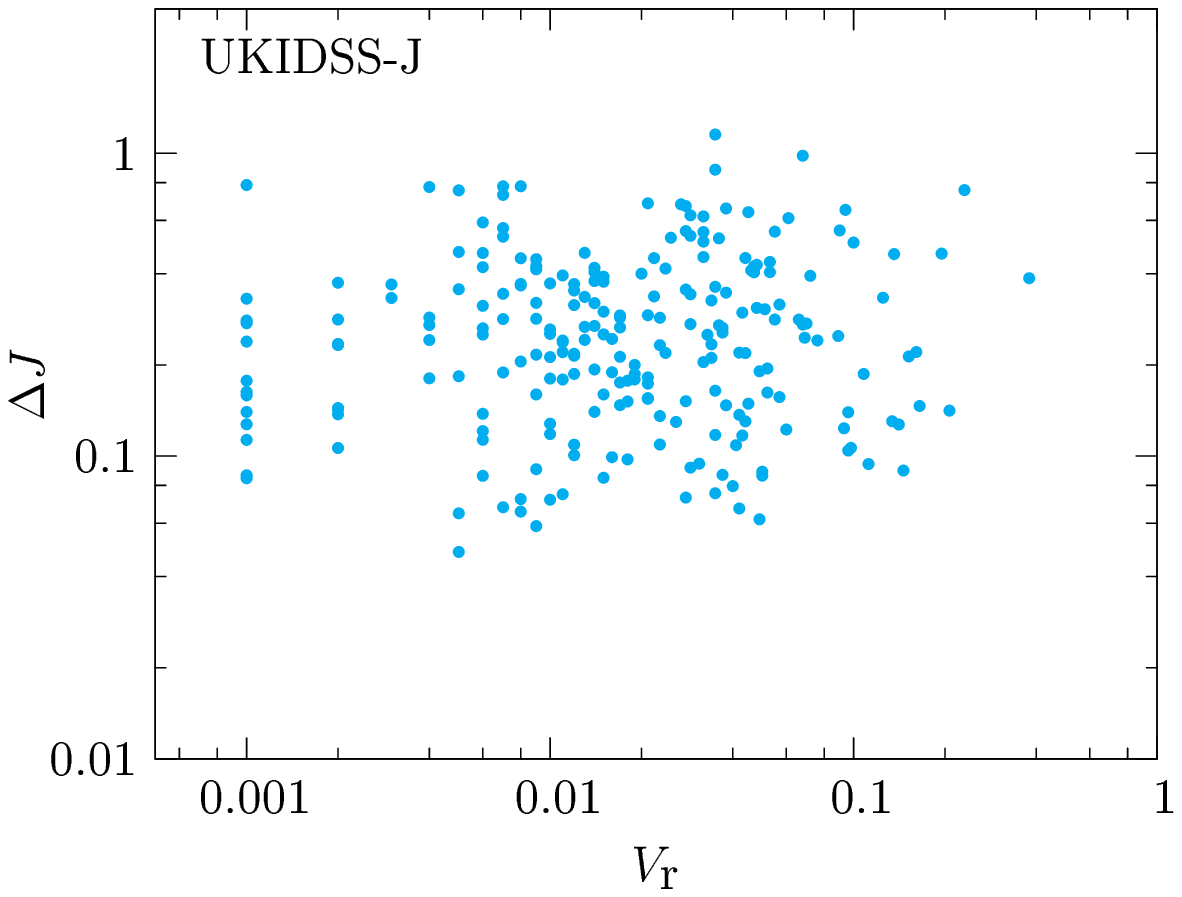}}
		\resizebox{75mm}{!}{\includegraphics[clip]{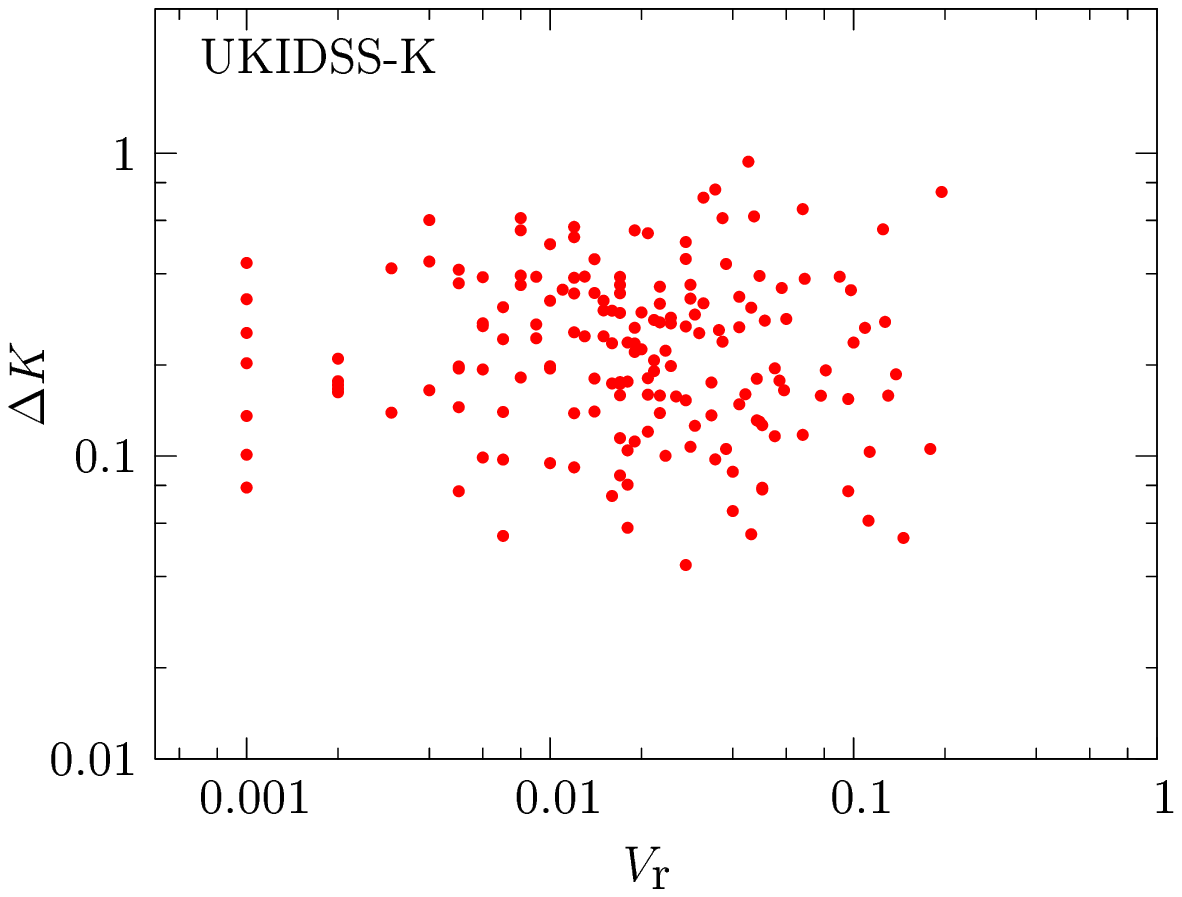}}
	\end{center}
		\caption{Optical variability (SDSS r-band) versus near-infrared variability amplitude. 
				The optical variability for each object is taken from \citet{Meusinger2011-AA}. 
  \label{Vr-Variability}}
\end{figure*}

%%%%%%%%%%%%%%%%%%%%%%%%%%%%%%
\subsection{Structure Function}\label{Structure-Function}
The ensemble variability as a function of rest-frame time lag is identical to the structure function (SF). 
The SF has been used to study both individual quasars and quasar ensembles and 
is useful for understanding the timescales of the relevant variability processes. 

The logarithmic slope ($\gamma$) of the SF can be an indicator for discriminating physical models and 
is described as follows: 
\begin{equation}
V \propto \tau^\gamma, 
\end{equation}
where $\tau$ is the rest-frame time lag with the time dilation effect caused by cosmic expansion, taken into account. 
The rest-frame time lag is given by $\tau=\Delta t/(1+z)$, 
where $\Delta t$ is the time lag in the observer-frame. 
\citet{Kawaguchi1998-ApJ} derived $\gamma$ values by constructing theoretical models. 
Their results are $\gamma \sim 0.74$--$0.90$ in the starburst model and $\gamma \sim 0.41$--$0.49$ in the disk instability model. 
Additionally, by using a microlensing model, \citet{Hawkins2002-MNRAS} predicted $\gamma = 0.25 \pm 0.03$. 

Figure \ref{SF} shows the SF for each sample. 
As done in Section \ref{Absolute-Magnitude}, 
the samples are divided into subsets according to the redshift. 
In the optical, the ensemble variability tends to increase with increasing rest-frame time lag 
for the UKIDSS-J and UKIDSS-K samples. 
The logarithmic slopes of these samples are computed by weighted least square fits 
to be $\gamma=0.13$ (UKIDSS-J) and $\gamma=0.66$ (UKIDSS-K), respectively. 
Several studies have shown that the SF in the optical increases steadily for $\tau \lesssim 2$--$3$ yr 
and becomes relatively constant for larger time lags \citep{Hook1994-MNRAS,Trevese1994-ApJ,Cristiani1996-AA}. 
Accordingly, we attribute the more-level slope in the UKIDSS-J sample to the difference in time ranges. 
In other words, the slope of the SF of the UKIDSS-J sample is expected to be less than that of the UKIDSS-K sample 
because a majority of the data in the UKIDSS-J sample have time lags longer than $\tau=2$--$3$ yr 
whereas a majority of the data in the UKIDSS-K sample have time lags shorter than $\tau=2$--$3$ yr. 
Meanwhile, the results for the DENIS samples are inconsistent with those from both the UKIDSS samples and previous studies. 
Although the ensemble variability of the DENIS-J sample decreases with increasing rest-frame time lag, 
the number of objects in the DENIS-J sample is considerably smaller than that in the UKIDSS-J sample. 
The DENIS-K sample, with $z>1.6$, has a notably smaller number of objects than the other samples. 
Therefore, we expect the results derived from the DENIS samples to be less reliable than those derived from the UKIDSS samples. 

In contrast, in the near-infrared, the ensemble variability of all samples appears to be anticorrelated with the rest-frame time lag. 
The logarithmic slopes $\gamma$ are $-0.002$ (DENIS-J), $-0.82$ (UKIDSS-J), 
$-0.14$ (DENIS-K), and $-0.67$ (UKIDSS-K). 
Even the DENIS samples show negative correlations with the rest-frame time lag, 
despite their being relatively unreliable because of the same reason mentioned in the above. 
If these trends are real, 
then Figure \ref{SF} shows that the decrease is moderate at shorter time lags ($\tau \lesssim 1000$) 
whereas the decrease is steep at larger time lags ($\tau \gtrsim 1000$). 
We further discuss the near-infrared SFs in Section \ref{Discussion}.

\begin{figure*}[thbp]
	\begin{center}
		\resizebox{75mm}{!}{\includegraphics[clip]{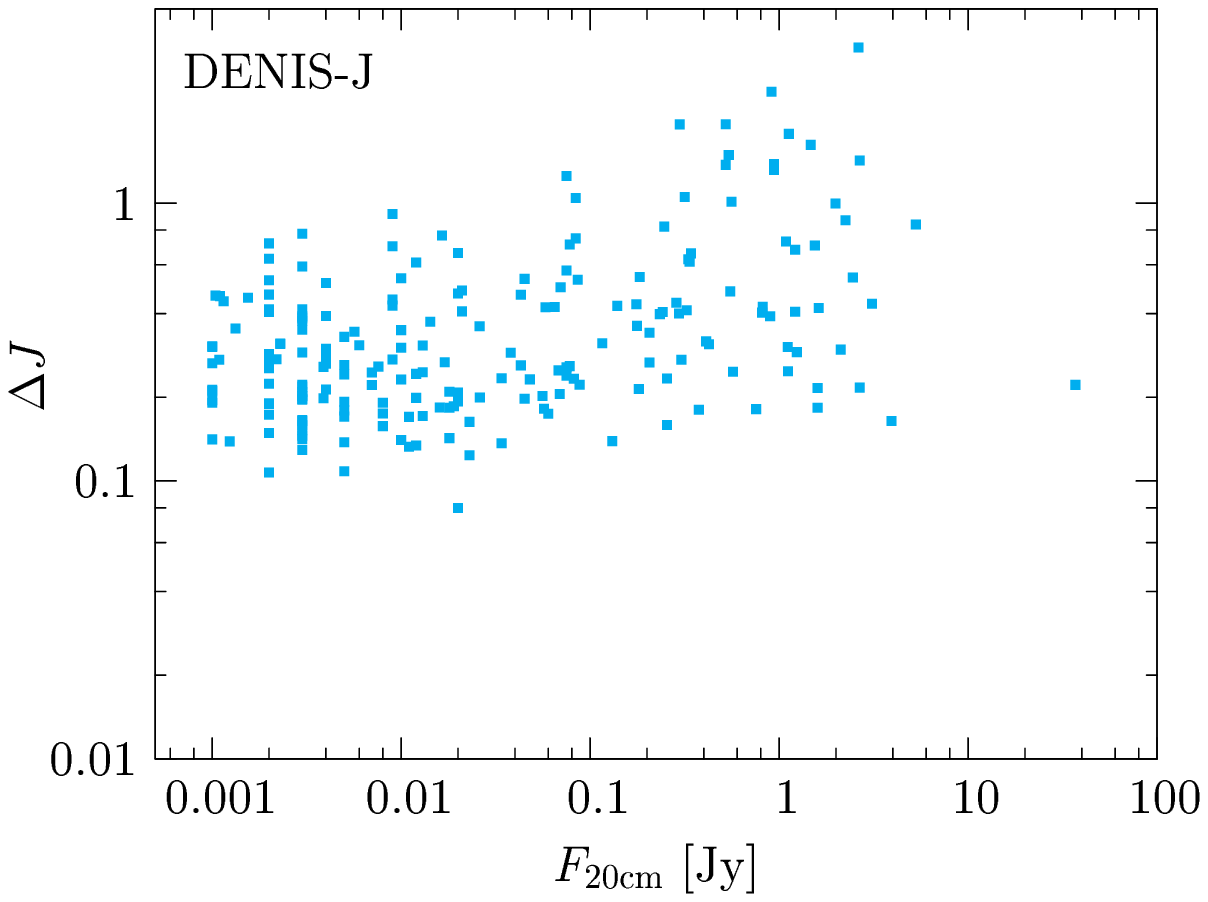}}
		\resizebox{75mm}{!}{\includegraphics[clip]{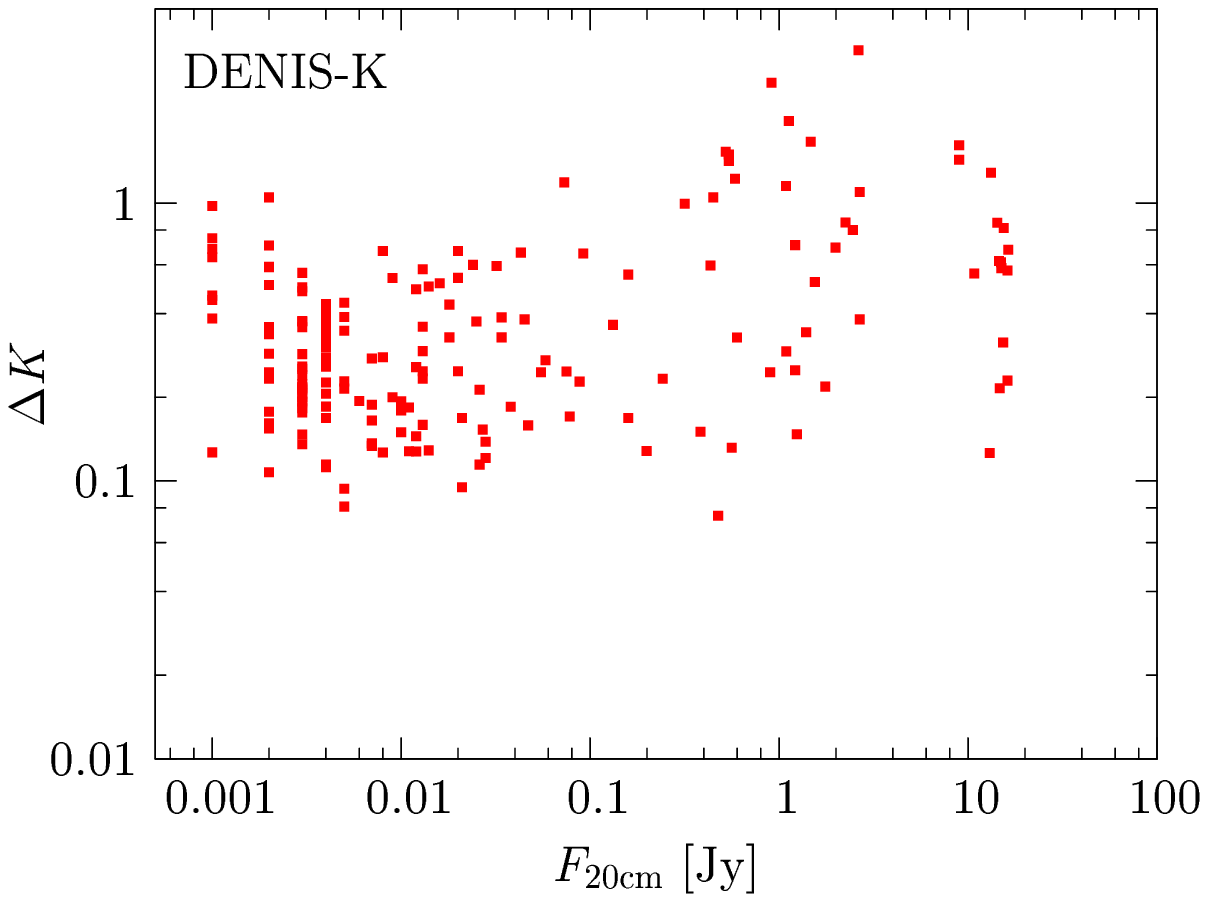}}
		\resizebox{75mm}{!}{\includegraphics[clip]{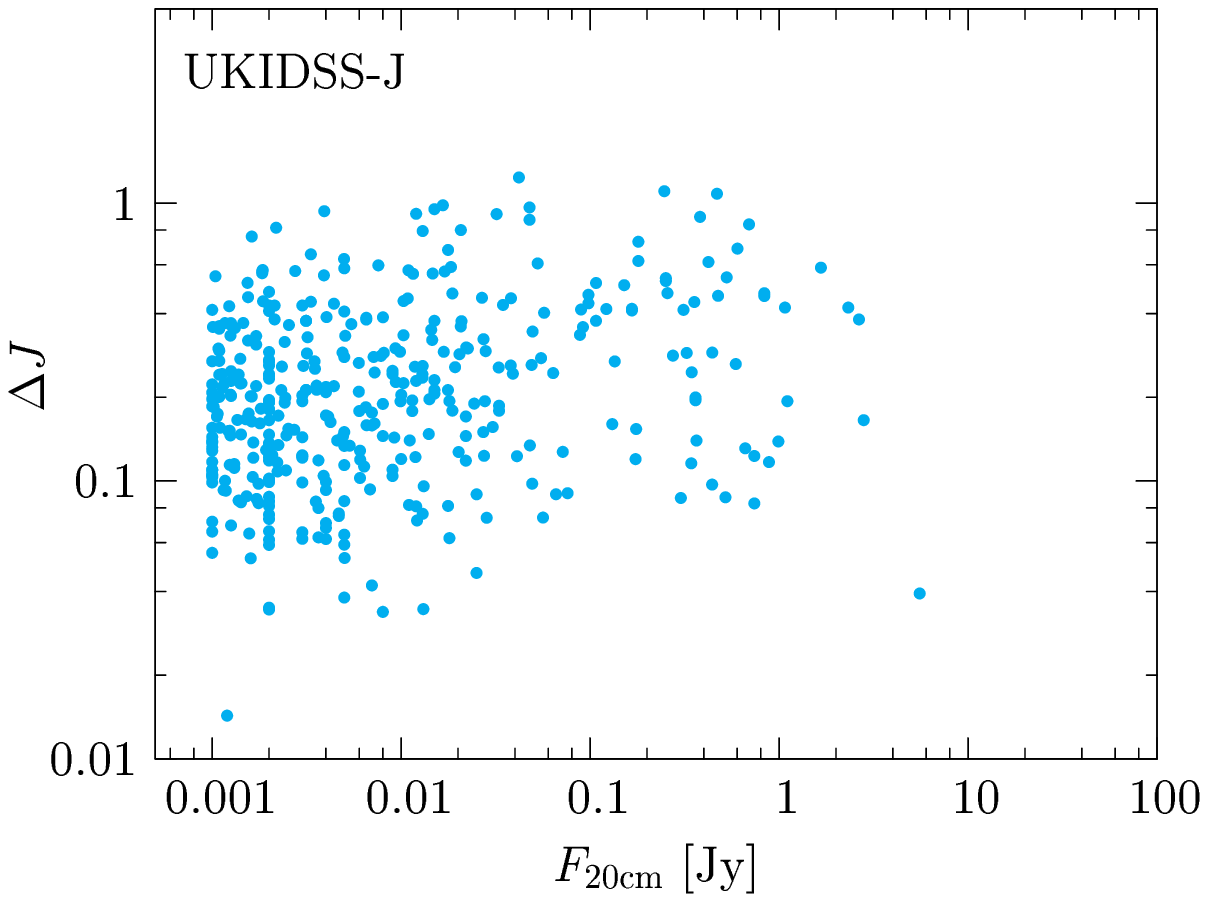}}
		\resizebox{75mm}{!}{\includegraphics[clip]{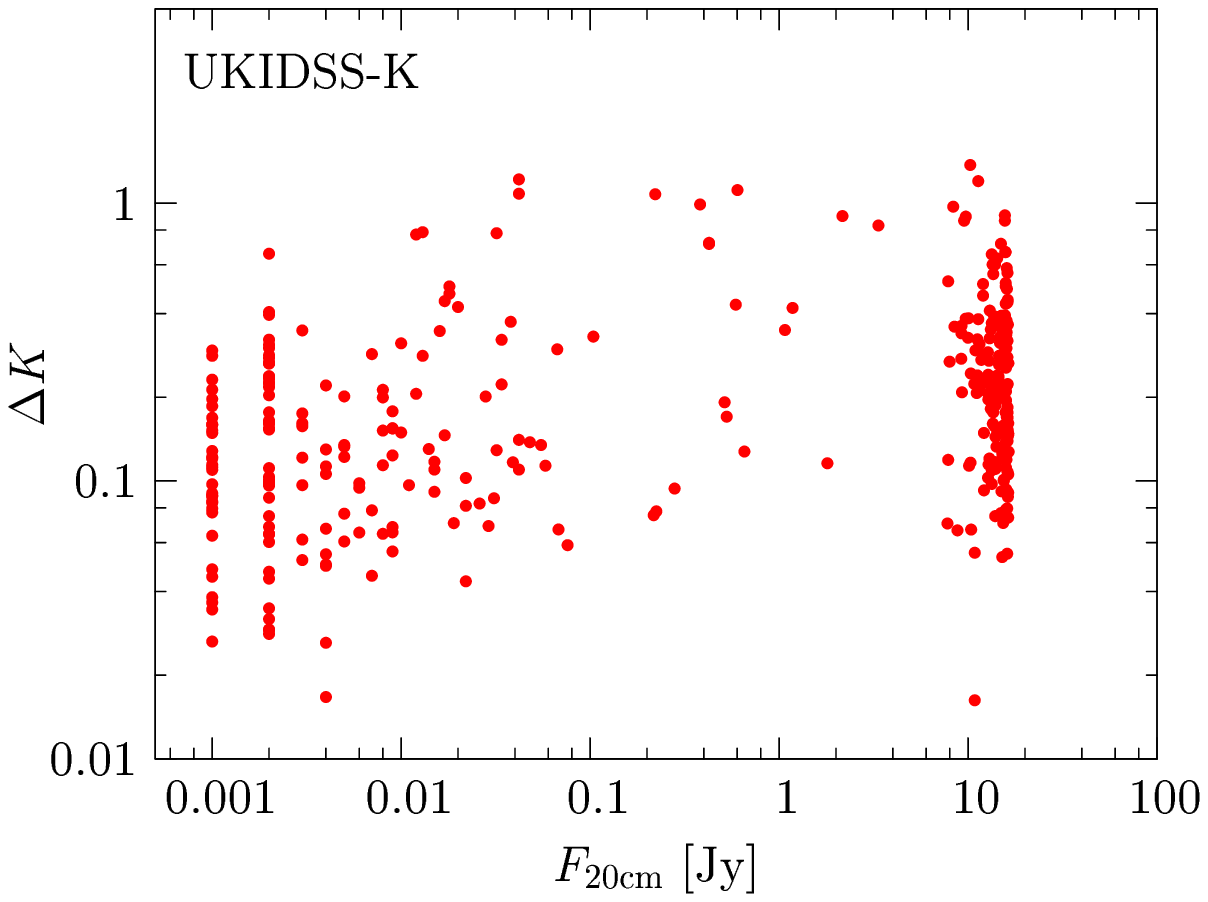}}
	\end{center}
		\caption{Radio flux versus near-infrared variability amplitude. 
				Radio flux values taken from either the NVSS or FIRST survey. 
  \label{F20-Variability}}
\end{figure*}

%%%%%%%%%%%%%%%%%%%%%%%%%%%%%%
\subsection{Multiwavelength Properties}
\subsubsection{Optical Variability}
\citet{Meusinger2011-AA} studied the UV/optical ensemble properties of over 9,000 quasars from the SDSS Stripe 82, 
and confirmed the well-known dependence of optical variability on time lag, luminosity, wavelength, and redshift. 
The complete table of their quasar sample is available at the CDS, 
which contains optical variability represented by the variability indicator in their paper. 
Using their variability indicators, we examine here a correlation between optical variability and near-infrared variability. 

We first cross-identified our sample objects with their sample quasars and then extracted optical variability for each cross-identified object. 
We cross-identified 25 objects for DENIS-J, 238 for UKIDSS-J, 22 for DENIS-K, and 173 for UKIDSS-K. 
Figure \ref{Vr-Variability} shows the optical variability at the SDSS r-band versus near-infrared variability amplitude. 
There seem to be positive correlations in the J-selected samples but negative correlations in the K-selected samples. 
The Spearman rank-correlation coefficients are 0.113 (DENIS-J), 0.058 (UKIDSS-J), $-0.107$ (DENIS-K), and $-0.076$ (UKIDSS-K) 
with the significant at $41\%$, $62\%$, $36\%$, and $68\%$ confidence levels, respectively. 
Although these levels are insufficient, 
there might be a weak correlation in each sample. 
Note that the correlations for the J-selected samples are opposite those for the K-selected samples. 

Because the near-infrared variability of our samples is simply the magnitude difference between double-epoch magnitudes, 
it does not directly represent the characteristic amplitude of the intrinsic variability. 
In addition, there are a statistically small number of objects in the samples. 
Consequently, the correlations must be confirmed by using a sample with a larger number of objects.

\begin{figure*}[thbp]
	\begin{center}
		\resizebox{75mm}{!}{\includegraphics[clip]{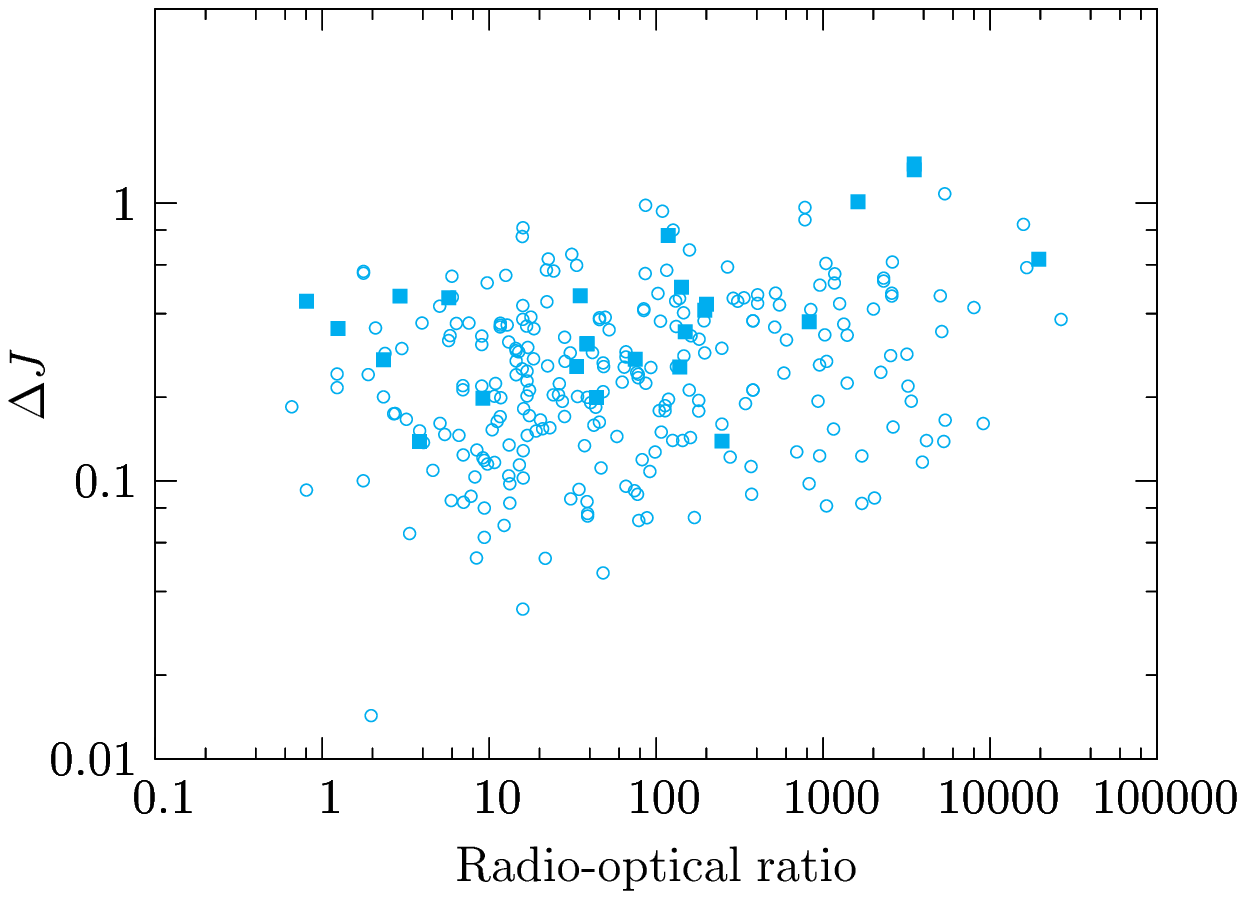}}
		\resizebox{75mm}{!}{\includegraphics[clip]{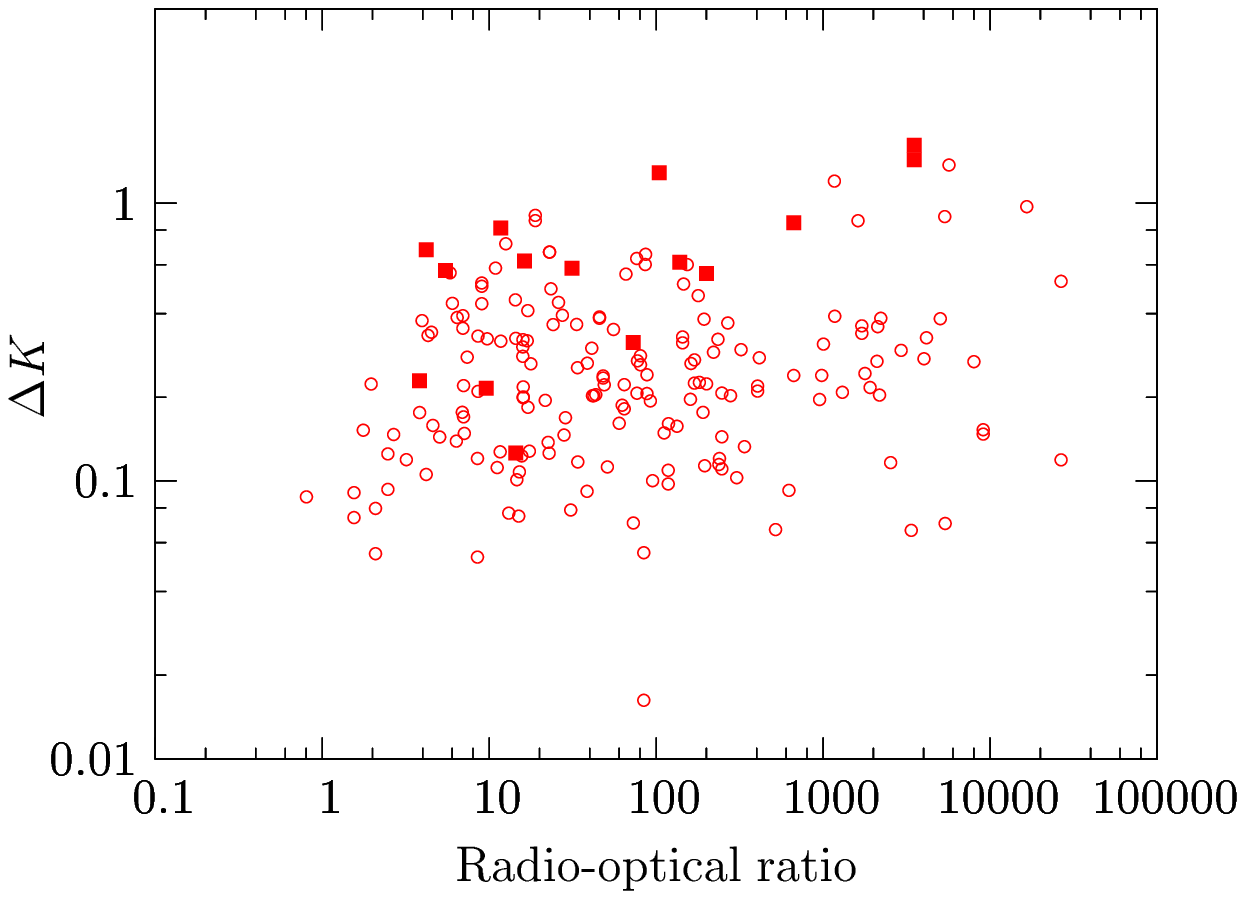}}
	\end{center}
		\caption{Radio-to-optical flux ratio ($F_\textnormal{\tiny 20 cm}/F_\textnormal{u}$) 
				versus near-infrared variability amplitude for variable AGNs selected from the SQ catalog. 
				Both radio flux at 20 cm and optical flux at the SDSS u-band are taken from the SQ catalog. 
				The left (right) panel shows the J-selected (K-selected) sample selected from the SQ catalog. 
				The solid squares and open circles represent the DENIS and UKIDSS samples, respectively. 
  \label{F20Fu-Variability}}
\end{figure*}

%%%%%%%%%%%%%%%%%%%%
\subsubsection{Radio}
Several studies have reported that radio-loud quasars tend to have larger optical variability amplitude than radio-quiet quasars 
\citep{Pica1983-ApJ,Garcia1999-MNRAS}. 
However, some recent studies have shown that there is no significant dependence of optical variability amplitude on radio loudness 
\citep{Helfand2001-AJ,Bauer2009-ApJ}. 
\citet{VandenBerk2004-ApJ} report that radio-loud quasars are more optically variable than radio-quiet quasars 
whereas there is no significant difference in optical variability between radio-detected and radio-undetected quasars. 
These trends were confirmed by \citet{Rengstorf2006-AJ} who used 933 known quasars in the QUEST Variability Survey. 

The SQ and QA catalogs contain radio flux extracted from 
the Faint Images of the Radio Sky at Twenty-Centimeters (FIRST) survey 
and either the National Radio Astronomy Observatory Very Large Array Sky Survey (NVSS) or FIRST surveys, respectively. 
DENIS-J has 209 objects with radio flux data, UKIDSS-J has 404, DENIS-K has 179, and UKIDSS-K has 353. 

We now review the relations between radio intensity and near-infrared variability. 
Figure \ref{F20-Variability} shows radio flux at 20 cm versus near-infrared variability amplitude. 
The radio flux is represented by the unit Jy. 
Because the SQ catalog provides 20 cm radio flux in the form of AB magnitude, 
we converted the radio magnitudes into radio flux in the unit of Jy. 
Each sample shows a weak correlation; 
that is, the near-infrared variability gradually increases with increasing radio intensity. 
The Spearman rank-correlation coefficients for the samples are 
0.346 (DENIS-J), 0.242 (UKIDSS-J), 0.235 (DENIS-K), and 0.270 (UKIDSS-K). 
Each of these values has a significance level below $1\%$ under the null hypothesis that 
variability and radio intensity are not correlated. 
Note that we ruled out the sources having $F_{\textnormal{\tiny 20 cm}} \gtrsim 10$, 
because they appear to be saturated as seen in the figure for the UKIDSS-K sample. 

We notice that positive correlations appear to be stronger above 0.01--0.1 Jy. 
Several studies have reported that radio-loud quasars tend to be more variable than radio-quiet quasars, 
although the optical variability of simple radio-detected quasars is not significantly 
different from that of radio-undetected quasars \citep{VandenBerk2004-ApJ,Rengstorf2006-AJ}. 
Our samples confirm these trends in the correlation between radio intensity and near-infrared variability amplitude. 

Moreover, there appears to be a positive correlation between radio-loudness and near-infrared variability. 
Figure \ref{F20Fu-Variability} shows the radio-to-optical flux ratio versus near-infrared variability amplitude. 
We define here the radio-to-optical ratio as the ratio of radio flux at 20 cm to the optical flux at the SDSS u-band. 
Both radio and optical fluxes are taken from the SQ catalog. 
We find positive correlations in Figure \ref{F20Fu-Variability}, which are significant 
because the Spearman rank-correlation coefficients are 0.407 (DENIS-J), 0.230 (UKIDSS-J), 0.570 (DENIS-K), and 0.134 (UKIDSS-K) 
with the significant at 96\%, 99\%, 97\%, and 92\% confidence levels, respectively. 

Theoretical calculations for near-infrared variability
of AGNs predict larger variability amplitude for orientations closer to the pole-on view \citep{Kawaguchi2011-pre}.
If the radio emission of our AGN sample originates in (short or small) jets, then the radio intensity
is likely to be correlated with the viewing angle in the sense that intense emission 
is associated to nearly pole-on view.
Thus, the correlation between the near-infrared variability amplitude and the radio intensity 
that we discovered in this study is consistent with this prediction.

%%%%%%%%%%%%%%%%%%%%%%%%%%%%%%%%%%%%%%%%%%%%%%%%%%
\section{Discussion}\label{Discussion}
\subsection{Optical Variability}
The intrinsic optical variability in both the DENIS and UKIDSS samples is inconsistent with 
the well-known negative correlation between rest-frame wavelength and variability. 
As mentioned in Section \ref{Rest-frame-Wavelength}, this is probably due to a biased sampling of high-redshift AGNs. 
However, the optical variability of the UKIDSS samples exhibit positive correlation with both the absolute magnitude and rest-frame time lag, 
which is consistent with previous analyses. 
On the other hand, although the absolute magnitude of the DENIS samples exhibit positive correlation with optical variability, 
the SF appears to be inconsistent with previous analyses. 
Therefore, compared with the DENIS samples, the UKIDSS samples strongly reflect the intrinsic properties of the optical variability for AGNs. 

We attribute the result that the variability properties of the UKIDSS samples are more consistent than those of the DENIS samples 
to the following reasons: 
First, the UKIDSS samples contain significantly more number of objects (over three times more; see Table \ref{Variable-Number}) 
than the DENIS samples contain, and a statistically small number of objects increases errors. 
Second, photometric errors of the UKIDSS samples are roughly five times lesser than those of the DENIS samples. 
Consequently, both statistically and photometrically, the UKIDSS samples should be more reliable than the DENIS samples.

%%%%%%%%%%%%%%%%%%%%%%%%%%%%%%
\subsection{Near-infrared Variability}
The near-infrared variability is anticorrelated with rest-frame wavelength, 
which is consistent with previous suggestions. 
Meanwhile, the near-infrared variability is anticorrelated with both absolute magnitude and rest-frame time lag,
which are the opposite of these correlations of the optical variability. 
As mentioned in Section \ref{Absolute-Magnitude}, a subset of these correlations are consistent with the radio-loud AGNs reported by \citet{Enya2002-ApJS}. 
Despite the different optical properties between the DENIS and UKIDSS samples, 
both show similar near-infrared variability properties. 

We focus here on the near-infrared SFs of our samples. 
The difference between the optical SF and the near-infrared SF is evident from Figure \ref{SF}. 
Contrary to the optical SFs, the near-infrared SFs exhibit negative correlations. 
In particular, the UKIDSS samples show strong anticorrelations over $\sim$1000 days. 
Note that the first-order SF is related to the autocorrelation function as follows \citep{Simonetti1985-ApJ}: 
\begin{equation}
V(\tau)=2 \sigma^2 [1-\rho (\tau)], 
\end{equation}
where $\sigma$ is the variance of a process and $\rho(\tau)$ is the autocorrelation function with $\rho(0) \equiv 1$. 
In this scheme, a negative SF slope indicates that the autocorrelation function increases with increasing rest-frame time lag. 
Accordingly, our samples imply that the near-infrared variability exhibits more periodic fluctuations on a longer time scale (up to $\sim$10 yr). 

This anticorrelation can be seen in the SFs of several quasars in \citet{Neugebauer1989-AJ} and \citet{Neugebauer1999-AJ}. 
Several quasars in their samples show SFs that decrease over a period of 3--5 yr. 
In particular, the SF for quasar PG 1226+023 is apparent. 
Our UKIDSS samples are consistent with this property. 
\citet{Neugebauer1999-AJ} also insist that there is ``periodic'' (sinusoidal-like) behavior 
over nearly a decade in the J, H, and K bands and 
suggest that this periodic behavior is a feature of the nonthermal engine of the radio-loud quasars. 
The peak near 3--5 yr that is seen in quasar PG 1226+023 appears to be found in the UKIDSS-K sample. 
Therefore, the SFs for the UKIDSS samples also show a periodic property similar to that found for radio-loud quasars, 
which is consistent with the analyses in Sections \ref{Redshift} and \ref{Absolute-Magnitude}.

%%%%%%%%%%%%%%%%%%%%%%%%%%%%%%%%%%%%%%%%%%%%%%%%%%
\section{Conclusion}
We have examined the ensemble variability of over 5,000 near-infrared luminous AGNs 
extracted from two AGN catalogs. 
The sample AGNs are divided into subsets according to whether, in the rest-frame, 
the near-infrared emission or the optical emission is at the origin of the near-infrared light, and 
we have demonstrated the dependence of intrinsic optical/near-infrared variability on rest-frame wavelength, redshift, 
absolute magnitude, and rest-frame time lag (i.e., structure function). 
The correlations of the optical variability with absolute magnitude and rest-frame time lag are qualitatively consistent 
with previous analyses, 
whereas the optical variability is inconsistent with the well-known negative correlation of optical variability with rest-frame wavelength. 
The inconsistencies in the optical variability properties reported herein and those found in previous analyses are probably 
due to a biased sampling of high-redshift AGNs. 
Furthermore, compared to the DENIS samples, the UKIDSS samples reflect known properties of the optical variability. 

However, the near-infrared variability is negatively correlated with the rest-frame wavelength, 
as has been suggested by previous studies. 
Nevertheless, it is notable that the near-infrared variability is anticorrelated with both the absolute magnitude and rest-frame time lag, 
which are opposite to the correlations with optical variability. 
The anticorrelation with absolute magnitude can be explained by the near-infrared emission from tori 
by using the theoretical model of \citet{Kawaguchi2011-pre}. 
The anticorrelation in the SFs should indicate that the near-infrared variability is more periodic over a longer time scale 
(up to $\sim$10 yr). 
The detailed physical process behind the near-infrared variability needs to be revealed. 
By comparing the properties of our samples with the properties of samples from \citet{Enya2002-ApJS} and \citet{Neugebauer1999-AJ}, 
we find that the general properties of the near-infrared ensemble variability 
are similar to the properties of radio-loud quasars. 

In addition, we examined the dependence of the near-infrared variability amplitude on 
optical variability, radio intensity, and radio-to-optical flux ratio. 
Although a subset of our samples indicates that the optical variability is 
positively (negatively) correlated with variability amplitude at the J (K) band, 
these correlations might not be significant. 
The near-infrared variability amplitude tends to increase with increasing radio intensity and radio-to-optical flux ratio, and 
this positive correlation with radio intensity tends to be stronger in the range of $F_{\textnormal{\tiny 20 cm}}>0.01$--0.1 Jy.

\begin{acknowledgements}
This publication makes use of data products from the Two Micron All Sky Survey, 
which is a joint project of the University of Massachusetts and 
the Infrared Processing and Analysis Center/California Institute of Technology, 
funded by the National Aeronautics and Space Administration and the National Science Foundation. 
SK is grateful to Toshihiro Kawaguchi for useful discussions. 
We thank the anonymous referee for helpful comments. 
\end{acknowledgements}

\end{document}